\documentclass[12pt]{article}
\usepackage{amsfonts}
\usepackage{amsmath, amsthm}
\usepackage{epsfig}
\usepackage{algorithm}

\usepackage{soul}
\usepackage[normalem]{ulem}
\usepackage{amsfonts}
\usepackage{subcaption}
\usepackage{epsfig}
\usepackage{booktabs}
\usepackage{lscape}
\usepackage{multirow}
\usepackage{longtable}
\usepackage{rotating}
\usepackage{array}
\usepackage{multicol}
\usepackage{graphicx}
\usepackage{bigstrut}
\usepackage{setspace}
\usepackage{epstopdf}
\usepackage{mathrsfs}
\usepackage{amsfonts}
\usepackage{epsfig}
\usepackage{booktabs}
\usepackage{threeparttable}
\usepackage{epstopdf}
\usepackage{lscape}
\usepackage{multirow}

\usepackage{algorithmic}
\usepackage{amsmath,verbatim,color,amssymb,epsfig}
\usepackage{bm}
\usepackage{amsfonts}
\usepackage{epsfig}
\usepackage{multirow}
\usepackage{graphicx}
\usepackage{array}
\usepackage[table]{xcolor}
\definecolor{Gray}{gray}{0.80}
\usepackage{lscape}
\usepackage{natbib}
\usepackage[normalem]{ulem}
\usepackage[colorlinks=true,urlcolor=blue,citecolor=black,linkcolor=blue,bookmarks=true]{hyperref}
\usepackage{caption}
\begin{document}
\def\eqx"#1"{{\label{#1}}}
\def\eqn"#1"{{\ref{#1}}}

\makeatletter % make @ act like a letter
\@addtoreset{equation}{section}
\makeatother  % make @ act like a non-letter

\def\yincomment#1{\vskip 2mm\boxit{\vskip 2mm{\color{red}\bf#1} {\color{blue}\bf --Yin\vskip 2mm}}\vskip 2mm}
\def\squarebox#1{\hbox to #1{\hfill\vbox to #1{\vfill}}}
\def\boxit#1{\vbox{\hrule\hbox{\vrule\kern6pt
          \vbox{\kern6pt#1\kern6pt}\kern6pt\vrule}\hrule}}

\newcommand{\blue}[1]{\textcolor{blue}{{#1}}}
\newcommand{\red}[1]{\textcolor{red}{{#1}}}
\def\theequation{\thesection.\arabic{equation}}
\newcommand{\ds}{\displaystyle}
\newtheorem{Pro}{Proposition}

\newcommand{\bJ}{\mbox{\bf J}}
\newcommand{\bF}{\mbox{\bf F}}
\newcommand{\bM}{\mbox{\bf M}}
\newcommand{\bR}{\mbox{\bf R}}
\newcommand{\bZ}{\mbox{\bf Z}}
\newcommand{\bX}{\mbox{\bf X}}
\newcommand{\bx}{\mbox{\bf x}}
\newcommand{\bQ}{\mbox{\bf Q}}
\newcommand{\bH}{\mbox{\bf H}}
\newcommand{\bh}{\mbox{\bf h}}
\newcommand{\bz}{\mbox{\bf z}}
\newcommand{\ba}{\mbox{\bf a}}
\newcommand{\bG}{\mbox{\bf G}}
\newcommand{\bB}{\mbox{\bf B}}
\newcommand{\bb}{\mbox{\bf b}}
\newcommand{\bA}{\mbox{\bf A}}
\newcommand{\bC}{\mbox{\bf C}}
\newcommand{\bI}{\mbox{\bf I}}
\newcommand{\bD}{\mbox{\bf D}}
\newcommand{\bU}{\mbox{\bf U}}
\newcommand{\bc}{\mbox{\bf c}}
\newcommand{\bd}{\mbox{\bf d}}
\newcommand{\bs}{\mbox{\bf s}}
\newcommand{\bS}{\mbox{\bf S}}
\newcommand{\bV}{\mbox{\bf V}}
\newcommand{\bv}{\mbox{\bf v}}
\newcommand{\bW}{\mbox{\bf W}}
\newcommand{\bw}{\mbox{\bf w}}
\newcommand{\bg}{\mbox{\bf g}}
\newcommand{\bu}{\mbox{\bf u}}

\newcommand{\bcU}{\boldsymbol{\cal U}}
\newcommand{\bbeta}{\boldsymbol{\beta}}
\newcommand{\bdelta}{\boldsymbol{\delta}}
\newcommand{\bDelta}{\boldsymbol{\Delta}}
\newcommand{\boldeta}{\boldsymbol{\eta}}
\newcommand{\bxi}{\boldsymbol{\xi}}
\newcommand{\bGamma}{\boldsymbol{\Gamma}}
\newcommand{\bSigma}{\boldsymbol{\Sigma}}
\newcommand{\balpha}{\boldsymbol{\alpha}}
\newcommand{\bOmega}{\boldsymbol{\Omega}}
\newcommand{\btheta}{\boldsymbol{\theta}}
\newcommand{\bmu}{\boldsymbol{\mu}}
\newcommand{\bnu}{\boldsymbol{\nu}}
\newcommand{\bgamma}{\boldsymbol{\gamma}}

\newcommand{\bse}{\begin{eqnarray*}}
\newcommand{\ese}{\end{eqnarray*}}
\newcommand{\be}{\begin{eqnarray}}
\newcommand{\ee}{\end{eqnarray}}
\newcommand{\bsq}{\begin{equation*}}
\newcommand{\esq}{\end{equation*}}
\newcommand{\bq}{\begin{equation}}
\newcommand{\eq}{\end{equation}}
\newcommand{\var}{\mbox{var}}
\newcommand{\trace}{\hbox{trace}}
\newcommand{\wh}{\widehat}
\newcommand{\wt}{\widetilde}
\newcommand{\eff}{_{\rm eff}}
\newcommand{\sub}{{\rm sub}}
\newcommand{\cat}{{\rm cat}}
\newcommand{\eLL}{\mathcal L}
\newcommand{\n}{\nonumber}
\newcommand{\bias}{\mbox{bias}}
\newcommand{\vecl}{\mbox{vecl}}
\newcommand{\AIC}{\mbox{AIC}}
\newcommand{\BIC}{\mbox{BIC}}
\newcommand{\MSE}{\mbox{MSE}}
\newcommand{\rank}{\mbox{rank}}
\newcommand{\cov}{\mbox{cov}}
\newcommand{\corr}{\mbox{corr}}
\newcommand{\argmin}{\mbox{argmin}}
\newcommand{\argmax}{\mbox{argmax}}
\newcommand{\diag}{\mbox{diag}}
\newcommand{\trans}{^{\rm \top}}
\newcommand{\bTheta}{\boldsymbol\Theta}
\newcommand{\bta}{\boldsymbol\eta}
\newcommand{\bphi}{\boldsymbol\phi}
\newcommand{\btau}{\boldsymbol\tau}
\newcommand{\boeta}{\boldsymbol\eta}
\newcommand{\bpsi}{\boldsymbol\psi}
\newcommand{\0}{{\bf 0}}
\newcommand{\A}{{\bf A}}
\newcommand{\U}{{\bf U}}
\newcommand{\V}{{\bf V}}
\newcommand{\e}{{\bf e}}
\newcommand{\R}{{\bf R}}
\newcommand{\G}{{\bf G}}
\newcommand{\bO}{{\bf O}}
\newcommand{\B}{{\bf B}}
\newcommand{\D}{{\bf D}}
\newcommand{\K}{{\bf K}}
\newcommand{\g}{{\bf g}}
\newcommand{\f}{{\bf f}}
\newcommand{\h}{{\bf h}}
\newcommand{\I}{{\bf I}}
\newcommand{\M}{\mbox{ $\mathcal{M}$}}
\newcommand{\BB}{\mbox{ $\mathcal{B}$}}
\newcommand{\N}{\mbox{ $\mathcal{N}$}}
\newcommand{\T}{{\bf T}}
\newcommand{\bP}{{\bf P}}
\newcommand{\s}{{\bf s}}
\newcommand{\m}{{\bf m}}
\newcommand{\W}{{\bf W}}
\newcommand{\w}{{\bf w}}
\newcommand{\X}{{\bf X}}
\newcommand{\x}{{\bf x}}
\newcommand{\tx}{{\widetilde \x}}
\newcommand{\Y}{{\bf Y}}
\newcommand{\C}{{\bf C}}
\newcommand{\tY}{{\widetilde Y}}
\newcommand{\y}{{\bf y}}
\newcommand{\Z}{{\bf Z}}
\newcommand{\z}{{\bf z}}
\newcommand{\Ybar}{{\overline{Y}}}
\newcommand{\Xbar}{{\overline{\X}}}
\newcommand{\xbar}{{\overline{\x}}}
\newcommand{\wbar}{{\overline{\W}}}
\newcommand{\bSig}{{\bf \Sigma}}
\newcommand{\bLam}{{\bf \Lambda}}
\def\th{^{th}}
\def\S{{\bf S}}
\def\L{{\bf L}}
\def\u{{\bf u}}
\def\v{{\bf v}}
\def\T{{\bf T}}
\def\bO{{\bf O}}
\def\I{{\bf I}}
\def\K{{\bf K}}
\def\t{{\bf t}}
\def\b{{\bf b}}
\def\r{{\bf r}}
\def\V{{\bf V}}
\def\c{{\bf c}}
\def\a{{\bf a}}
\def\vec{\mbox{vec}}

\newcommand{\cvec}[1]{{\mathbf #1}}
\newcommand{\rvec}[1]{\vec{\mathbf #1}}
\newcommand{\minor}{{\rm minor}}
\newcommand{\spn}{{\rm Span}}
\newcommand{\range}{{\rm range}}
\newcommand{\mdiv}{{\rm div}}
\newcommand{\proj}{{\rm proj}}
\newcommand{\RR}{\mathbb{R}}
\newcommand{\NN}{\mathbb{N}}
\newcommand{\QQ}{\mathbb{Q}}
\newcommand{\ZZ}{\mathbb{Z}}
\newcommand{\EE}{\mathbb{E}}
\newcommand{\<}{\langle}
\renewcommand{\>}{\rangle}
\renewcommand{\emptyset}{\varnothing}
\newcommand{\attn}[1]{\textbf{#1}}
\newcommand{\bproof}{\bigskip {\bf Proof. }}
\newcommand{\eproof}{\hfill\qedsymbol}
\newcommand{\Disp}{\displaystyle}
\newcommand{\qe}{\hfill\(\bigtriangledown\)}
\newcommand*{\dif}{\mathop{}\!\mathrm{d}}

\newtheorem{thm}{Theorem}[section]
\newtheorem{lem}{Lemma}[section]
\newtheorem{rem}{Remark}[section]
\newtheorem{cor}{Corollary}[section]
\newcolumntype{L}[1]{>{\raggedright\let\newline\\\arraybackslash\hspace{0pt}}m{#1}}
\newcolumntype{C}[1]{>{\centering\let\newline\\\arraybackslash\hspace{0pt}}m{#1}}
\newcolumntype{R}[1]{>{\raggedleft\let\newline\\\arraybackslash\hspace{0pt}}m{#1}}

\newcommand{\tabincell}[2]{\begin{tabular}{@{}#1@{}}#2\end{tabular}}

\newtheorem{theorem}{Theorem}
\newtheorem{definition}{Definition}

\newcommand{\dT}{\top}

\newcommand{\algorithmicobs}{\textbf{Observations:}}
\newcommand{\algorithmicprior}{\textbf{Prior:}}
\newcommand{\PRIOR}{\item[\algorithmicprior]}
\newcommand{\OBS}{\item[\algorithmicobs]}

\newcommand{\algorithmicoutput}{\textbf{Output:}}
\newcommand{\OUTPUT}{\item[\algorithmicoutput]}

\setcitestyle{authoryear,open={(},close={)}}

\baselineskip=24pt
\begin{center}
{\Large \bf
Oncology Dose Finding Using Approximate Bayesian Computation Design
}
\end{center}

\vspace{2mm}
\begin{center}
{\bf Huaqing Jin, Wenbin Du and Guosheng Yin$^{*}$ }
\end{center}

\begin{center}

Department of Statistics and Actuarial Science\\
The University of Hong Kong\\
Pokfulam Road, Hong Kong\\

\vspace{2mm}

%haoluns@sfu.ca
{\em *Correspondence}: gyin@hku.hk\\

\end{center}
\noindent{\bf Abstract}\\
In the development of new cancer treatment, an essential step is to determine the maximum tolerated dose (MTD) via phase I clinical trials.
Generally speaking, phase I trial designs can be classified as either model-based or algorithm-based approaches.
Model-based phase I designs are typically more efficient by using all observed data,
while there is a potential risk of model misspecification that may lead to unreliable dose assignment and incorrect MTD identification. In contrast,
most of the algorithm-based designs are less efficient in using cumulative information, because they tend to focus on the observed data in the neighborhood of the current dose level for dose movement. 
To use the data more efficiently yet without any model assumption,
we propose a novel approximate Bayesian computation (ABC) approach for phase I trial design.
Not only is the ABC design free of any dose--toxicity curve assumption, but it can also aggregate all the available information accrued in the trial for dose assignment.
Extensive simulation studies demonstrate its robustness and efficiency compared with other phase I designs. 
We apply the ABC design to the MEK inhibitor selumetinib trial to demonstrate its satisfactory performance.
The proposed design can be a useful addition to the family of phase I clinical trial designs due to its simplicity, efficiency and robustness.

\vspace{0.5cm}

\vspace{1cm}
	\section{Introduction}
	In oncology research, a phase I clinical trial often aims to determine 
the maximum tolerated dose (MTD),
which is typically defined as the dose with the dose-limiting toxicity (DLT) probability closest to the target toxicity rate \citep{yin2012clinical}.
In the development of new drugs, 
phase I clinical trials play an essential role
because the selected MTD will 
be further investigated in the subsequent phase II or III trials.
Misidentification of the MTD would lead to 
an unreliable conclusion and may even cause termination of a trial immaturely. As a result, this leads to a waste of abundant resources
or risking patients treated at the excessively toxic doses,
which violates the ethical principle. 
Moreover, as the first-in-human study, 
subjects available for a phase I trial
are rather limited, and the sample size is typically around $35$ \citep{iasonos2014adaptive},
and thus it is challenging to identify the MTD
with such a small sample size.

Various phase I trial designs have been proposed with the goal to
determine the MTD both efficiently and accurately.
Depending on whether to adopt a model assumption on the dose--toxicity curve or not, 
these designs can generally be classified into two branches:
the algorithm-based (model-free or curve-free) and the model-based (often under a parametric modeling assumption) approaches.
Among the algorithm-based designs, the $3+3$ design \citep{storer1989design} is the most commonly used one for phase I oncology trials due to its simplicity and conservativeness \citep{yuan2011bayesian}.
%The popularity of the $3+3$ design is due to its simplicity and clarity. 
However, it has also been criticized for the poor performance due to inefficient use of the data. 
%Thus, after $3+3$ designs, 
Alternatively, a variety of model-free designs have been developed to improve the trial efficiency in determining the MTD.
Gasparini and Eisele \citep{gasparini2000curve} presented a curve-free method in which the probabilities of toxicity are directly modeled as an unknown multidimensional parameter.
Ivanova et al. \citep{ivanova2007cumulative} proposed a cumulative cohort design (CCD) where
the dose escalation or de-escalation criterion is based on the Markov chain theory.
Adopting a beta--binomial Bayesian model and a probabilistic up-and-down rule,
Ji et al. \citep{ji2007dose, ji2007bayesian} developed an interval design where all possible dose
assignment actions can be tabulated in a spreadsheet.
This method is further improved 
%with a new statistic, 
by using a unit probability mass
to enhance the safety, which is called the modified toxicity probability interval (mTPI) design \citep{ji2010modified}.
Under the Bayesian framework, Liu and Yuan \citep{liu2015bayesian} proposed the 
Bayesian optimal interval (BOIN) design by minimizing the probability of incorrect dose allocation.
By refining the mTPI design, Yan et al. \citep{yan2017keyboard} proposed the keyboard design 
%by partitioning the toxicity probability scale into 
using more and shorter toxicity probability intervals.
Enlightened by the uniformly most powerful Bayesian test \citep{johnson2013uniformly}, Lin and Yin \citep{lin2018uniformly} developed the uniformly most powerful Bayesian interval
(UMPBI) design for phase I dose-finding trials.

Along the line of directly modeling the dose--toxicity curve, many model-based phase I designs have been proposed.
The continual reassessment method (CRM) \citep{o1990continual, o1996continual}
is the most popular model-based design,
which often uses a single unknown parameter to link the true toxicity probabilities with
the prespecified toxicity probabilities of dose levels. %Following the CRM method, 
Cheung and Chappell \citep{cheung2000sequential} extended the CRM by incorporating weights to account for the late-onset toxicity outcomes.
Another extension of the CRM focuses on modeling bivariate
competing outcomes \citep{braun2002bivariate}.
Incorporated with the Bayesian model averaging approach, Yin and Yuan \citep{yin2009bayesian} developed the Bayesian model averaging CRM to 
overcome the arbitrariness of skeleton specification (i.e., the prespecified toxicity probabilities of all doses) and thus enhance the model robustness.
Other extensions of the CRM include O'Quigley and Paoletti \citep{o2003continual}; Yuan et al. \citep{yuan2007continual}; Wages et al. \citep{wages2011continual}.
%Another well-known model-based design is 
For safety concerns, the escalation with overdose control (EWOC)
\citep{babb1998cancer} is designed to locate the MTD as quickly as possible subject to the constraint that the predicted proportion of patients receiving overdoses does not exceed a specified threshold. 

Our work is motivated by a phase I trial on the MEK inhibitor selumetinib in children with progressive low-grade gliomas (LGG) \citep{banerjee2017phase}.
There were three candidate doses in the trial: 25, 33, and 43 mg/$\mathrm{m}^2$/dose bis in die.
The target DLT rate was $\phi=0.25$. Originally, the trial adopted 
the likelihood-based modified CRM using a 2-parameter logistic model based on dosages adjusted for the body surface area with an initial cohort size of 3 patients at each new dose level.
However, as the prior information on the dose--toxicity curve is very limited at the
phase I trial stage, 
the model-based designs might be at risk of violating the parametric assumption,
which undermines the efficiency of the dose-finding procedure. Moreover, with only three doses under investigation, a parametric model may not fit the data well.
On the other hand, if we choose an algorithm-based method to select a dose for an incoming cohort of patients, 
most of the existing methods only consider the data 
%from the patients treated 
collected at the current dose level,
which weakens the efficiency of the design without usage of all the cumulative data in the trial thus far. 
For example, 
a typical interval design determines the next dose level solely based on the data from the current dose level.
To alleviate the risk of model misspecification 
and utilize all the 
available information accrued in the trial,
we develop an approximate Bayesian computation (ABC) design 
to identify the MTD.
The ABC design adopts the idea from the approximate Bayesian computation sampling methods \citep{rubin1984bayesianly} 
and generates the weighted posterior samples based on all the available data
%for each cohort of patients
without any complex formulas and dose--toxicity model assumptions.
Based on the weighted samples, decisions
among dose escalation, de-escalation, or retaining the current dose
are made for each cohort.
In this way, 
our method avoids introducing any explicit dose--toxicity model, yet 
can utilize the cumulative information from all dose levels 
when selecting the next dose level. 
Extensive simulation studies demonstrate that the ABC design
is efficient compared with the state-of-the-art methods, while 
it is robust due to its model-free or curve-free feature.

The rest of the paper is organized as follows.
In Section 2, we introduce the ABC 
design for dose finding in phase I clinical trials.
We present the simulation studies to
evaluate the operating characteristics of the new method and compare the ABC design with
several well-known phase I designs in Section 3. 
An application to the phase I trial of the MEK inhibitor selumetinib is provided in Section 4. 
The paper is concluded with a brief discussion in Section~5.
\section{Methodology} \label{sec_metho}
Suppose that a phase I clinical trial aims to investigate $K$ dose levels with 
the corresponding DLT rates $p_1 < \cdots < p_K$.
The target toxicity rate is denoted as $\phi$.
After enrolling and treating the first $n$ cohorts of patients,  
the current dose level is denoted by $d_n$. Thus far, we observe the cumulative data $Y_n=\{y_k\}_{k=1}^K$ and $M_n=\{m_k\}_{k=1}^K$,
where $y_k$ represents the number of observed DLTs and $m_k$ represents the number of 
patients treated at dose level $k$.

The main goal of the ABC design is to adopt all the available information when deciding the dose level for the next cohort without imposing any explicit dose--toxicity model assumptions. 
In the Bayesian framework, 
%provides an natural and elegant way to incorporate the collected data for dose escalation.
we first assign a prior $\pi(p_1, \ldots, p_K)$ on $\{p_k\}_{k=1}^K$, and then 
we can update the posterior distribution of $(p_1, \ldots, p_K)$ as 
\bse
\pi(p_1, \ldots, p_K|Y_n, M_n) \propto \pi(p_1, \ldots, p_K) P(Y_n|\{p_k\}_{k=1}^K, M_n),
\ese
where $P(Y_n|\{p_k\}_{k=1}^K, M_n)$ is the likelihood function.
Based on the posterior distribution, the next dose level can be selected efficiently.
The major difficulty lies on how to choose a suitable and simple prior $\pi(p_1, \ldots, p_K)$
while taking the monotonicity constraint $p_1 < \cdots < p_K$ into account.
The dilemma is that a parametric model assumption typically undermines the robustness of the design, while due to the monotonicity constraint, 
the specific form of the prior $\pi(p_1, \ldots, p_K)$ without such a model assumption can be complicated. The unusual complexity of the prior would also diminishes the flexibility of the design.
Further, it causes the difficulty on calculating the posterior distribution as well as the subsequent Bayesian inference.
%One of the main advantages of the Bayesian method is that it allows to select different priors under different contexts, which ensures the efficient usage of the data.
%If the prior is complicated, the change of the prior will cause a lot of tedious work which is not desirable in practice. 

\subsection{Approximate Bayesian Computation}
To circumvent the aforementioned problem, 
we use the idea from the approximate Bayesian computation (ABC) sampling methods \citep{rubin1984bayesianly}.
The ABC methods are typically used to handle the problem that for some complex models, 
the evaluation of the likelihood function is computationally costly or elusive. 
In such cases, the Bayesian inference with the closed form of the posterior distribution is prohibitive. 
The ABC methods bypass the evaluation of the likelihood function with Monte Carlo simulations. 

The most naive ABC method is known as the ABC rejection algorithm \citep{beaumont2010approximate}.
Suppose that we are interested in obtaining the posterior samples of parameter $\theta$ given the dataset $D$ and prior $\pi(\theta)$.
We can carry out the ABC rejection algorithm as follows, 
\begin{enumerate}
    \item Draw a sample $\hat{\theta}$ from the prior $\pi(\theta)$.
    \item Given the sample point $\hat{\theta}$, generate a dataset $\wh{D}$ under the likelihood model $P(D|\theta)$, i.e., using the prior predictive distribution.
    \item If the generated dataset $\wh{D}$ is close to the observed dataset $D$ under some prespecified distance measure, we keep the sample $\hat{\theta}$ as the posterior sample; otherwise, the sample point is discarded or rejected.
    \item The procedure is repeated for a large number of times to obtain the adequate posterior samples.
\end{enumerate}

In the phase I trial under our setting, the evaluation of the likelihood is rather simple, 
while the prior model sometimes can be too complicated to give a closed form 
due to the monotonicity constraint.
On the other hand, the monotonicity constraint is natural when we implement Monte Carlo simulations.
For example, we can first generate the samples without any constraints and then sort them in an
ascending order to obtain the prior samples of $\{p_k\}_{k=1}^K$.
Thus, the ABC methods provide a very simple solution to obtain the posterior samples of $\{p_{k}\}_{k=1}^K$
in a Monte Carlo manner.

For illustration of the monotonically constrained sampling problem, we show an example on how to sample 
%under the monotonicity constraint naturally. We consider 
three variables $\{X_i\}_{i=1}^3$ whose base distribution is 
$\mathrm{Uniform}(0, 1)$ but with the constraint $0< X_1 < X_2 < X_3 < 1$.
\begin{Pro}
The probability density function (pdf) of the joint uniform distribution of $(X_1, X_2, X_3)$ under the constraint $0< X_1 < X_2 < X_3 < 1$ is 
\be\label{toy}
f(x_1, x_2, x_3) = f_1(x_1|x_2) f_2(x_2) f_3(x_3|x_2),
\ee
where 
\bse
f_2(x_2) \sim \mathrm{Beta}(2, 2), \  
f_1(x_1|x_2) \sim \mathrm{Uniform}(0, x_2), \ 
f_3(x_3|x_2) \sim \mathrm{Uniform}(x_2, 1).
\ese
\end{Pro}
\noindent\textbf{Proof:} By definition, the pdf of $(X_1, X_2, X_3)$ can be written as 
\bse
f(x_1, x_2, x_3) = C I_{\{0<x_1<x_2<x_3<1\}},
\ese
where $C$ is a constant and $I$ is an indicator function.
Noting that the joint pdf can be rewritten as
\bse
f(x_1, x_2, x_3) = 
C I_{\{0<x_1<1\}}
I_{\{0<x_2<1\}}
I_{\{0<x_3<1\}}
I_{\{x_1<x_2\}}
I_{\{x_2<x_3\}},
\ese
the marginal distribution of $X_2$ is obtained by
\bse
f_2(x_2)
&=&
\int \int f(x_1, x_2, x_3) \dif x_1 \dif x_3 \\
&=&
\int CI_{\{0\le x_3 \le 1\}}I_{\{0\le x_2 \le 1\}}I_{\{x_2<x_3\}}
\int I_{\{x_1<x_2\}}I_{\{0\le x_1 \le 1\}} \dif x_1 \dif x_3 \\
&=&
\int Cx_2I_{\{0\le x_3 \le 1\}}I_{\{0\le x_2 \le 1\}}I_{\{x_2<x_3\}}
\dif x_3 \\
&=&
C(1-x_2)x_2I_{\{0\le x_2 \le 1\}} \\
&=&
6(1-x_2)x_2I_{\{0\le x_2 \le 1\}},
\ese
which corresponds to the pdf of 
$\mathrm{Beta}(2, 2)$.
The conditional density of $(X_1, X_3)$ given $X_2$ is
\bse
f(x_1, x_3|x_2)
=
\frac{f(x_1, x_2, x_3)}{f_2(x_2)} 
=
\frac{I_{\{0\le x_1 \le x_2\}}I_{\{x_2 \le x_3 \le 1\}}}{(1-x_2)x_2} = f_1(x_1|x_2) f_3(x_3|x_2),
\ese
because given $X_2$, $X_1$ and $X_3$ are independent, and thus each follows a uniform distribution.
\qedsymbol

\begin{figure}[!ht]
    \centering
    \includegraphics[scale=0.4]{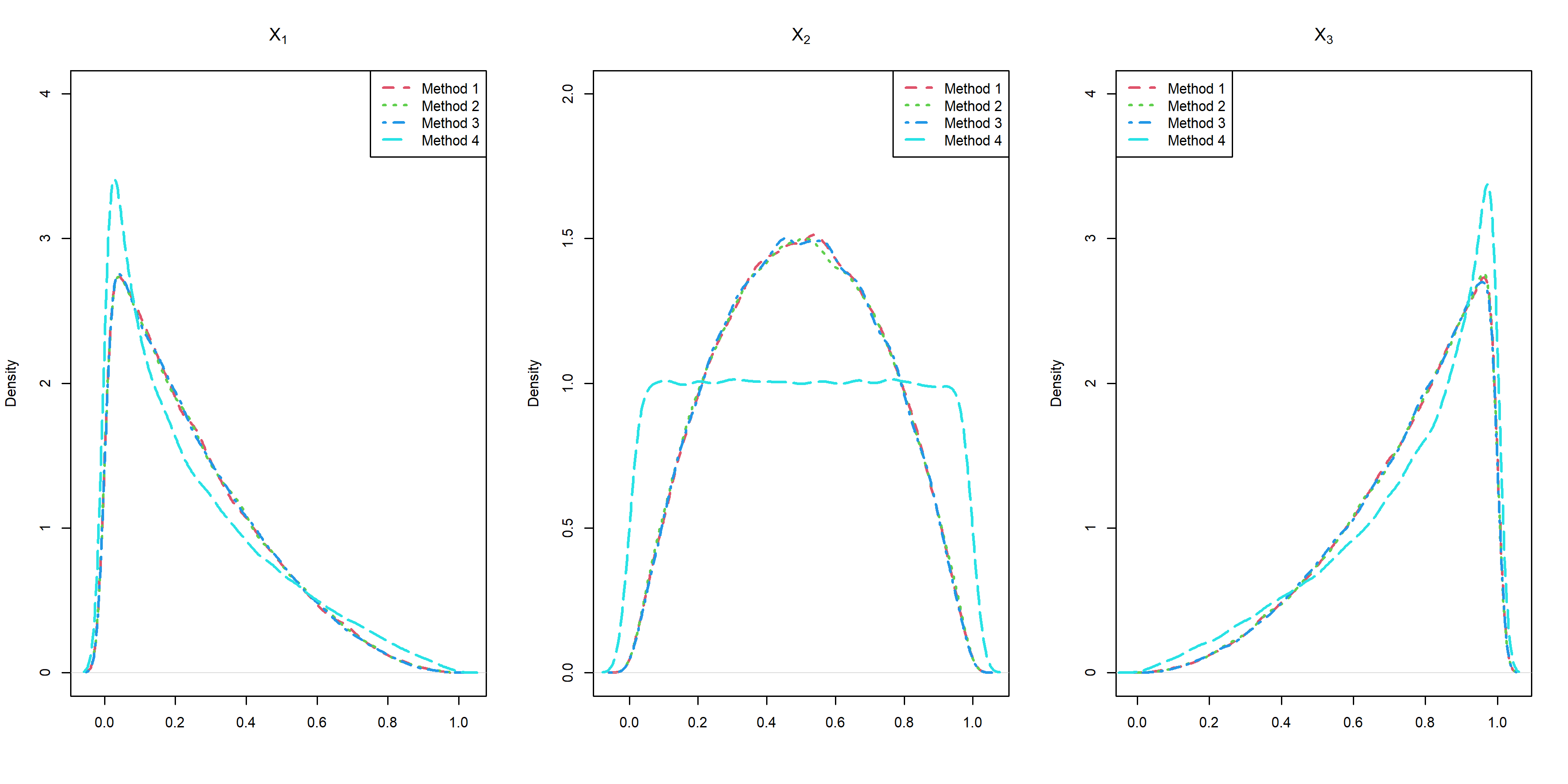}
    \caption{The densities of $X_1$ (left), $X_2$ (middle) and $X_3$ (right) under the four different sampling methods.}
    \label{fig:toy}
\end{figure}

We explore four ways to sample $(X_1, X_2, X_3)$ under the monotonicity constraint.
\begin{itemize}
    \item \textbf{Method 1:} sample $(X_1, X_2, X_3)$ following the distributions in~(\ref{toy}).
    \item \textbf{Method 2:} sample three variates from $\mathrm{Uniform}(0, 1)$ independently, and then sort them in an ascending order and label the sorted samples as $X_1, X_2, X_3$.
    \item \textbf{Method 3:} sample $X_1, X_2, X_3 \sim  \mathrm{Uniform}(0, 1)$ independently and only keep the samples satisfying $X_1 < X_2 < X_3$.
    \item \textbf{Method 4:} sample $(X_1, X_2, X_3)$ as follows, 
\bse
 X_2 &\sim& \mathrm{Uniform}(0, 1), \\ 
X_1|X_2 &\sim& \mathrm{Uniform}(0, X_2), \\
X_3|X_2 &\sim& \mathrm{Uniform}(X_2, 1).
\ese
\end{itemize}
The estimated densities of $(X_1, X_2, X_3)$ under the four sampling methods are shown in 
Figure~\ref{fig:toy}.
Clearly, the first three sampling methods lead to equivalent distributions while the last one yields a different one.
Method 1 requires tedious derivations which would be more complicated if more random variables are involved. 
Method 2 is natural and simple to incorporate the monotonicity constraint, and Method 3 is less efficient.

There are many extensions beyond the basic ABC rejection method, e.g., the ABC-importance method and ABC-Markov chain Monte Carlo method \citep{beaumont2010approximate}. 
These extensions are not suitable in the context of the phase I trial, because they add additional complexity in the generation of prior samples. 
Thus, our subsequent discussions focus on the ABC rejection method.

\subsection{Optimal Dose Selection}
With the ABC rejection algorithm in hand, we are still not ready to obtain the posterior samples of $\{p_k\}_{k=1}^K$ for the phase I trial.
The main challenge lies in the low efficiency of the ABC rejection method, because
%When implementing the ABC rejection algorithm,
the acceptance rate can be very low which makes the whole procedure rather slow 
to obtain an adequate number of posterior samples for reliable and robust inference.
Moreover, in the ABC rejection method, all the $\hat{\theta}$'s resulting in inconsistent $\wh{D}$'s with the observed data $D$ are equally discarded. 
However, even the discarded $\hat{\theta}$'s may provide some useful information about the posterior distribution.
%in estimating the optimal dose.
For example, some generated data $\wh{D}$'s may be very different from the original data $D$, while some may only moderately deviate from $D$, and thus 
we should not treat the corresponding $\hat{\theta}$'s equally.

Considering the both issues, we propose to modify the ABC rejection algorithm for the phase I trial as follows. 
\begin{enumerate}
    \item Select a suitable prior $\pi(p_1, \ldots, p_K)$ and a distance measure $\rho_h(\cdot, \cdot)$ and set a
    large number $J$. 

    \item Generate the prior samples $\{p^{(j)}_k\}_{k=1}^K$ for $J$ times from $\pi(p_1, \ldots, p_K)$.
    \item Given the prior samples $\{p^{(j)}_k\}_{k=1}^K$, we generate the corresponding datasets, $Y_n^{(j)}=\{y_k^{(j)}\}_{k=1}^K$ with $y_k^{(j)} \sim \mathrm{Binom}(p_k^{(j)}, m_k)$.
    \item Given $\{Y_n^{(j)}\}_{j=1}^J$, the weight for each sample $\{p_{k}^{(j)}\}_{k=1}^K$ can be obtained as $w^{(j)}=\rho_h(Y_n^{(j)}, Y_n)$, where $Y_n=\{y_k\}_{k=1}^K$ is the observed data.
    \item Use the weighted samples $\{(w^{(j)}, \{p_{k}^{(j)}\}_{k=1}^K)\}_{j=1}^J$ as the posterior samples.
\end{enumerate}

Similar to a Gaussian kernel,
we choose the distance measure as 
\bse
\rho_h(Y_n^{(j)}, Y_n) = \exp
\left\{-  
\frac{\sum_{k=1}^K(y_k^{(j)}/m_k- y_k/m_k)^2}{h}
\right\},
\ese
where $h$ is the bandwidth and the selection of $h$ will be discussed in Sensitivity Analysis Section.

This weighted procedure overcomes the low acceptance rate problem of the ABC rejection method 
and adopts all the prior (weighted) samples for the posterior inference.
We then estimate the toxicity rate for dose level $k$ as 
\bse
\hat{p}_{n, k}  = S\left(\{p_k^{(j)}\}_{j=1}^J, \{w^{(j)}\}_{j=1}^J\right),
\ese
where $S(\cdot, \cdot)$ can be any function yielding reasonable estimator of $\hat{p}_{n, k}$.
In our manuscript, we take $S(\cdot, \cdot)$ as the weighted median. 
The weighted median function is the $50\%$ weighted percentile of $\{(w^{(j)}, p_k^{(j)})\}_{j=1}^J$ defined as follows, 
\begin{itemize}
    \item Sort $\{p_k^{(j)}\}_{j=1}^J$ in an ascending order to obtain 
    $\{p_{k, (j)}\}_{j=1}^J$ with the corresponding weights $\{w_{(j)}\}_{j=1}^J$.
    \item The weighted median is selected as $p_{k, (l)}$ satisfying
    \bse
    \frac
    {\sum_{j=1}^{l-1} w_{(j)}}
    {\sum_{j=1}^{J} w_{(j)}} \le \frac{1}{2} \ \ \ \mathrm{and} \ \ \
    \frac{\sum_{j=l+1}^{J} w_{(j)}}
    {\sum_{j=1}^{J} w_{(j)}} \le \frac{1}{2}.
    \ese
\end{itemize}
We adopt the weighted median  function because it gives more robust estimator compared with the weighted mean function.

The optimal dose for cohort $n+1$ based on the current data $Y_n$ and $M_n$ is 
\bse\label{optdose}
d_{n+1}^* = \argmin_{k=1, \ldots, K} \left|\hat{p}_{n, k}-\phi\right|.
\ese

\subsection{Prior Elicitation}
While the ABC design is flexible with respect to the choice of the prior $\pi(p_1, \ldots, p_K)$,
the prior samples $\{p_k^{(j)}\}_{k=1}^K$ play an important role 
%in the ABC design 
for efficiently selecting the MTD.
%It is worth noting that by choosing a suitable prior with the dose--toxicity model, 
%the ABC method is essentially the Bayesian CRM design.
%However, 
To enhance the robustness of the ABC design, we avoid imposing any explicit model assumptions.
An intuitive way to generate samples without the dose--toxicity model is to follow Method 2 in Section 2.1, i.e., first generate $K$ samples from $\mathrm{Uniform}(0, 1)$, and then sort them in an ascending order.
%This method fails to take the target rate $\phi$ into consideration, as a result, the prior samples are less informative and lead to poor performance.
%In fact, in our experiments, with such generating method, 
%most of the subjects are allocated to the first two dose levels and thus the whole design is too conservative.

Before conducting a phase I trial, we are only given the target toxicity rate $\phi$ as well as the toxicity probability monotone constraint.
%$0\le p_1 < \cdots < p_K \le 1$ before the trial.
%Thus, it is desirable to encode both two parts of information in the prior to obtain more informative samples.
%When we pay our attention to the target rate $\phi$, we notice that the main goal of the phase I design is not to estimate the toxicity rates accurately but 
The main goal of the phase I design is to choose the dose level $k$ whose DLT rate is closed to $\phi$.
%Thus, instead of generating the samples from the whole sample space, 
Because each dose is possibly the MTD, we can generate prior samples of $p_1, \ldots, p_K$ from $K+1$ possible models $\{\mathcal{M}_k\}_{k=0}^K$,
where $\mathcal{M}_k$ is the model that dose level $k$ is the MTD while $\mathcal{M}_0$ indicates that all the dose levels are overly toxic (no MTD). This way of incorporating $\phi$ into the model provides more informative prior samples for the ABC design.

Consequently, we propose to generate the prior samples as follows.
\begin{enumerate}
    \item Considering the trade-off between computation and performance, we generate $20000$ prior samples of $p_1, \ldots, p_K$ from each model $\mathcal{M}_k$, so the total number of prior samples is $J=20000(K+1)$. 
    These prior samples can be generated before the trial conduct and saved for repeated use.
    
    \item Given $\mathcal{M}_k$ with $k\neq0$,
    \begin{enumerate}
        \item Set dose level $k$ as the target with a DLT rate $p^{(j)}_k\sim \text{Uniform}(\phi-\delta, \phi+\delta)$ where $\delta$ is a small prespecified number.
        \item Independently generate $k-1$ samples from  $\text{Uniform}(0,\phi-\delta)$ and sort them in an ascending order to obtain $\{p_1^{(j)}, \ldots, p_{k-1}^{(j)}\}$.
        \item Independently generate $K-k$ samples from  $\text{Uniform}(\phi+\delta, 2\phi)$ and sort them in an ascending order to obtain $\{p_{k+1}^{(j)}, \ldots, p_{K}^{(j)}\}$.
    \end{enumerate}
    \item Under $\mathcal{M}_0$,
        we independently generate $K$ samples from  $\text{Uniform}(\phi+\delta, 2\phi)$ and sort them in an ascending order to obtain $\{p_{1}^{(j)}, \ldots, p_{K}^{(j)}\}$.
\end{enumerate}

The neighborhood parameter $\delta$ controls the distinguishability of the target dose level in the generated prior samples. 
A larger value of $\delta$ indicates that the MTD is easier to be determined in the prior samples, and vice versa.
In Sensitivity Analysis Section, %~\ref{subsec:robust}, 
we conduct extensive simulation studies to show the ABC design is robust to the selection of $\delta$,
and we recommend to choose $\delta =0.1$ as a default value in practice.

\subsection{Dose-finding Algorithm}
To ensure the safety and benefit for the patients, 
we further impose an early stopping criterion.
In the implementation of the ABC design, 
we terminate the trial when there is strong evidence indicating the lowest dose level is still overly toxic.
We assign a $\mathrm{Beta}(0.5, 0.5)$ prior distribution to the DLT rate $p_1$, and
if $\Pr(p_1>\phi|y_1, m_1 \ge 3) >0.95$, the trial will be terminated for safety.
%we eliminate dose level $k$ and all those above $k$ from the trial if 
%$\Pr(p_k>\phi|y_k, m_k \ge 3) >0.95$,

%Here, we abandon the early elimination rule, because we think most of the phase I designs have their own innate mechanism to avoid selecting over-toxic dose levels and to better take advantage of the property for the phase I design, it is not necessary to impose the early elimination rule.
%In fact, the CRM design\citep{zhou2018accuracy} also does not have early elimination rule.
%In our simulation studies and the real data application, we remove the early elimination rule for all the designs for a fair comparison.

%In summary, 
The dose-finding procedure of the ABC design is detailed as follows. 
\begin{enumerate}
    \item Treat the first cohort of patients at the lowest or the physician-specified dose level.
    %and set the initial admissible set as $\mathcal{A}=\{1, \ldots, K\}$. 
    \item After enrolling $n$ cohorts, select the optimal dose level 
    $
    d_{n+1}^* = \argmin_{k=1, \ldots, K} \left|\hat{p}_{n, k}-\phi\right|
        $ based on (\ref{optdose}). 
     
    %If $\Pr(p_{d_n}>\phi|y_{d_n}, m_{d_n}\ge 3) > 0.95$, update $\mathcal{A}$ as $\mathcal{A} = \mathcal{A}\cap \{1, \ldots, d_n-1\}$.
    
    \item According to the optimal dose level $d_{n+1}^*$, 
    \begin{enumerate}
        \item If $d_n > d_{n+1}^*$, then $d_{n+1}=d_n-1$.
        \item If $d_n = d_{n+1}^*$, then $d_{n+1}=d_n$.
        %\item If $d_n = d_{n+1}^*$, then $d_{n+1}=\max_{k \le d_n, k \in \mathcal{A}}(k)$.
        %\item If $d_n < d_{n+1}^*$, then $d_{n+1}=\max_{k \le d_n+1, k \in \mathcal{A}}(k)$.
        \item If $d_n < d_{n+1}^*$, then $d_{n+1}=d_n+1$.
    \end{enumerate}
    \item The trial can be either stopped after exhaustion of the maximum sample size,  or be terminated
early for safety if the lowest dose level is too toxic by the early stopping rule, $\Pr(p_1>\phi|y_1, m_1\ge 3) >0.95$.
\end{enumerate}

At the end of the trial, the observed dataset $(Y_N, M_N)$ is collected, where $N$ is the total number of cohorts. 
The MTD is estimated with another round of ABC simulation, i.e., $d^*$ is selected as 
\bse
d^* = \argmin_{k=1, \ldots, K} \left|\hat{p}_{N, k}-\phi\right|.
\ese

Note that in the dose-finding algorithm, we adopt the same early termination rule as those in most of the interval designs 
\citep{liu2015bayesian, lin2018uniformly}  for a fair comparison.
However, as our prior samples involve model $\mathcal{M}_0$ (i.e., all doses are overly toxic), 
it is possible to construct our own early termination rule as 
\bse
\frac{\sum_{j=1}^J w^{(j)} I(p_1^{(j)} > \phi)} 
{\sum_{j=1}^J w^{(j)}} > t,
\ese
where $t$ is the threshold value, e.g., $t=0.9$.
\section{Simulation Studies}\label{sec:simu}
\subsection{Sensitivity Analysis} \label{subsec:robust}

We investigate the effect of the parameters $\delta$ and $h$ on the performance of the ABC design with 
the analysis of variance (ANOVA) method \citep{cangul2009testing}.
We randomly generate dose--toxicity scenarios following the approach of Paoletti et al. \citep{paoletti2004design},
for which the detailed procedure is described in Section A.2 of the Appendix.
To conduct a comprehensive analysis,
we consider possible settings with four different influential factors of the phase I trials,
including the average probability difference $\Delta$ around the target in the randomly generated scenario, 
the number  of dose levels $K$, 
the sample size
as well as the target toxicity rate $\phi$.
The first three factors affect the difficulty of the MTD-identification task,
where larger $\Delta$, smaller $K$, and larger sample size would typically result 
in a higher MTD selection percentage.
The possible levels of the four factors are listed in Table~\ref{anova},
which yields $4\times 3 \times 8 \times 3  = 288$ different settings.
Under each setting, we investigate the performance of the ABC design via
$1000$ randomly generated scenarios when $\delta$ takes a value of $\{0, 0.05, 0.10, 0.15, 0.20 \}$ or $\delta$ is randomly selected from Uniform$(0, 0.2)$
and $h$ is selected from $\{0.1, 0.05, 0.01, 0.005\}$.

%\begin{table}[!h]
%	\caption{
%	The simulation factors that may affect the dose-finding performance of phase I
%trial design and the results of ANOVA in terms of the percentage of MTD selection. The ANOVA
%also includes all the pairwise interactions between the five simulation factors.
%}
%	\label{anova}
%	\centering
%    \resizebox{\linewidth}{!}{
%	\begin{tabular}{lcrrr}
%		\toprule
%		Factors &Levels of factor & DF & SS &  MSE\\
%		\midrule
%		Average probability difference around $\phi$ ($\Delta$) &\{0.05, 0.07, 0.10, 0.15\} &3& 65.35 & 21.78 \\
%		Bandwidth parameter $h$ & \{0.1, 0.05, 0.01, 0.005\}&4& 4.32 & 1.44\\
%		Sample size&$\{18, 24, \dots,60\}$ & 7& 5.58 & 0.80 \\
%		Number of dose levels $K$ &\{3, 5,  7\} & 2 & 0.81 & 0.40\\
%		Target toxicity probability $\phi$ &\{0.25, 0.30, 0.33\}&2& 0.34 & 0.17\\
%		Neighborhood parameter $\delta$ & \{0.0, 0.05, 0.10, 0.15, 0.20, random\}&5& 0.54 & 0.11\\
%		\midrule
%		Total variance &  & 6911 &  83.39 &  \\
%		\bottomrule
%		\multicolumn{5}{l}{DF: degree of freedom; SS: sum of squares; MSE: mean squared error (MSE$=$SS/DF)}
%	\end{tabular}
%	}
%\end{table}

After obtaining the percentage of MTD selection for each setting under different values of $(\delta, h)$,
we perform ANOVA with regard to these percentages using the simulation factors including all the pairwise interactions in Table~\ref{anova}.
In the ANOVA, we also regard the tuning parameters $\delta$ and $h$ as factors in the evaluation of dose-finding performance.
Thus, the degree of freedom of the total variance for ANOVA is $288\times 6 \times 4-1=6911$.
In terms of the mean squared error (MSE), 
among the six influential factors, 
the neighborhood parameter $\delta$ has the least effect on the performance of the ABC design.
In fact, the tuning parameter $\delta$ only accounts for $0.64\%$ $(0.54/83.39)$ of the MTD selection percentage variance,
which indicates the ABC design is robust to the choice of $\delta$.
The performance of the ABC design is more sensitive to the choice of the bandwidth parameter $h$, as
it is the second most influential factor on the MTD selection percentage among the six factors.

%\begin{figure}[t]
%	\centering
%	\includegraphics[width=0.8\columnwidth]{anova_noEli.png}
%	\caption{
%	The MTD selection percentage versus the neighborhood parameter $\delta$ (left) and 
%	bandwidth parameter $h$ (right)
%	under  the probability difference around the target $\Delta=\{0.05, 0.07, 0.10, 0.15\}$.
%	We set $\delta=\{0, 0.05, 0.1, 0.15, 0.2\}$ respectively and also consider $\delta$ randomly chosen from Uniform$(0, 0.2)$ and $h$ takes a value of $\{0.005, 0.01, 0.05, 0.1\}$.
%	}
%	\label{fig:delta}
%\end{figure}

\begin{figure}[t]
	\centering
	\includegraphics[width=0.8\columnwidth]{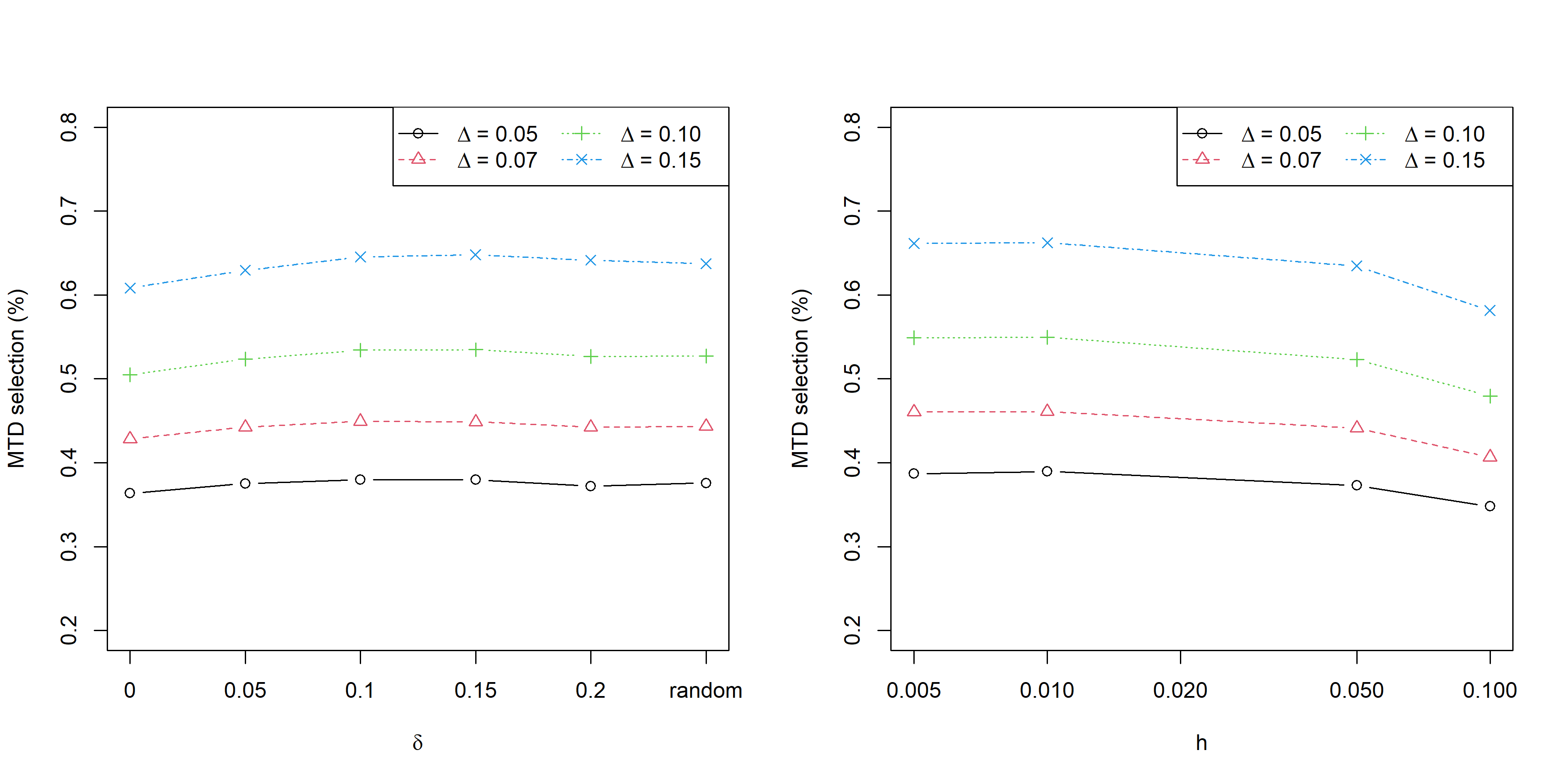}
	\caption{
	The MTD selection percentage versus the neighborhood parameter $\delta$ (left) and 
	bandwidth parameter $h$ (right)
	under  the probability difference around the target $\Delta=\{0.05, 0.07, 0.10, 0.15\}$.
	We set $\delta=\{0, 0.05, 0.1, 0.15, 0.2\}$ respectively and also consider $\delta$ randomly chosen from Uniform$(0, 0.2)$ and $h$ takes a value of $\{0.005, 0.01, 0.05, 0.1\}$.
	}
	\label{fig:delta}
\end{figure}

From Table~\ref{anova}, it is clear that the average probability difference $\Delta$ around the target is  
the dominating factor for the percentage of MTD selection,  
as it corresponds to the largest MSE (significantly larger than the one in the second place) in the ANOVA. 
We further study the effect of the tuning parameters $\delta$ and $h$ on the MTD selection percentage under different values of $\Delta$. 
The results are presented in Figure~\ref{fig:delta}, 
where we show the MTD selection percentage versus the tuning parameters $\delta$ and $h$ for 
$\Delta = \{0.05, 0.07, 0.10, 0.15\}$ respectively.
Under the settings with various levels of trial difficulty (the difficulty of MTD identification decreases as $\Delta$ increases),
the tuning parameter $\delta$ shows relatively minor effect on the performance of the ABC design, demonstrating the robustness of the design
and it is especially true when $\delta \in [0.05, 0.20]$.
As for the bandwidth parameter $h$, 
it is clear that a larger value of $h$ would undermine the performance of the ABC design.
When $h$ is decreased near $0.01$,
the performance of the ABC design is saturated and
further reduction of $h$ does not improve the performance significantly.
Thus, in practice, it is recommended to choose any value of $\delta \in [0.05, 0.20]$ 
and $h=0.01$ for the ABC design.
In the following simulation studies and real trial application, 
we set $\delta=0.1$ and $h=0.01$ throughout.

%\begin{figure}[h]
%	\begin{subfigure}[]{0.5\textwidth}
%		\centering
%		% include first image
%		\includegraphics[width=.9\linewidth]{ABC_random_c16_05.jpg}  
%		\caption{$\Delta=0.05$}
%	\end{subfigure}
%	\begin{subfigure}{0.5\textwidth}
%		\centering
%		% include second image
%		\includegraphics[width=.9\linewidth]{ABC_random_c16_07.jpg}
%		\subcaption{$\Delta=0.07$}
%	\end{subfigure}
%	
%	\begin{subfigure}[]{0.5\textwidth}
%		\centering
%		% include first image
%		\includegraphics[width=.9\linewidth]{ABC_random_c16_10.jpg}  
%		\caption{$\Delta=0.10$}
%	\end{subfigure}
%	\begin{subfigure}{0.5\textwidth}
%		\centering
%		% include second image
%		\includegraphics[width=.9\linewidth]{ABC_random_c16_15.jpg}
%		\subcaption{$\Delta=0.15$}
%	\end{subfigure}
%	%\newline
%	
%	\caption{Simulation results 
%	%for the dose-finding trials 
%	with sample size $30$ based on 5000 randomly generated
%		dose–toxicity scenarios under the average probability difference of $\Delta=0.05$, 0.07, 0.10 and 0.15
%		around the target toxicity probability $\phi= 0.30$, respectively.}
%	\label{fig:cpr30}
%\end{figure}

\subsection{Evaluation under Random Scenarios}\label{subsec:compare}
%We compare the ABC design with other phase I designs.
To make an extensive comparison of the ABC design with existing methods, 
we select six state-of-the-art phase I designs, including the CRM design with the power model whose model skeleton is selected
using the method of Lee and Cheung \citep{lee2009model},
the CCD design \citep{ivanova2007cumulative} which is based on the Markov chain theory, 
the modified toxicity probability interval (mTPI) design \citep{ji2010modified}, 
the Bayesian optimal interval design (BOIN) \citep{liu2015bayesian},
the keyboard design \citep{yan2017keyboard}, as well as the uniformly most powerful Bayesian interval (UMPBI) design \citep{lin2018uniformly}.
Among the six designs, 
the CRM is the only model-based one,
while the other five are algorithm-based methods and, more specifically, they are all interval designs.
Unless otherwise stated, 
all the six existing designs adopt the default parameters following the original papers
and we utilize the same early stopping rule for the five interval designs and the ABC design, i.e., 
%when $\Pr(p_k >\phi |y_k, m_k \ge 3) > 0.95$, we eliminate the corresponding dose level as well as those above $k$,
%and 
if $\Pr(p_1 >\phi |y_1, m_1 \ge 3) > 0.95$, we terminate the whole trial for safety.
As for the safety rule under the CRM design, the posterior probability of $p_1>\phi$ is calculated based on the CRM model and the threshold probability is still set at $0.95$.
The detailed settings for the six existing designs can be found in Section A.1 of the Appendix.

We set the target toxicity rate $\phi=0.30$ and 
investigate five dose levels for each trial. 
To assess the operating characteristics of the seven designs,
we conduct simulation studies under the sample size of $30$ and choose the cohort size as $3$.
To avoid cherry-picking cases, we randomly generate dose--toxicity scenarios following the approach of Paoletti et al. \citep{paoletti2004design},
for which the detailed generating method is given in Section A.2 of the Appendix.
The average probability difference $\Delta$ around  the target is controlled at 
$0.05$, $0.07$, $0.1$ and $0.15$ respectively, and under each value of $\Delta$, 
%average probability difference,
we replicate $5000$ simulations.

Four summary statistics are reported to evaluate the performances of the seven designs under comparison. 
The two main measurements, reflecting the accuracy and efficiency of a design, are
the percentage of MTD selection and the percentage of patients treated at the MTD (MTD allocation), for
which larger values are preferred.
The remaining two measurements quantify the safety aspects of a trial, including
the percentage of trials that select overdoses as the
MTD (overdose selection),
the percentage of patients allocated to overdoses (overdose allocation).
%the risk of high toxicity (which is defined as the
%percentage of trials that leads to the DLT rate greater than $\phi$),
%and the percentage of patients experiencing DLT.
A design with smaller values of these two safety statistics would be considered more ethical and desirable.

\begin{figure}[!ht]
	\begin{subfigure}[]{0.5\textwidth}
		\centering
		% include first image
		\includegraphics[width=.9\linewidth]{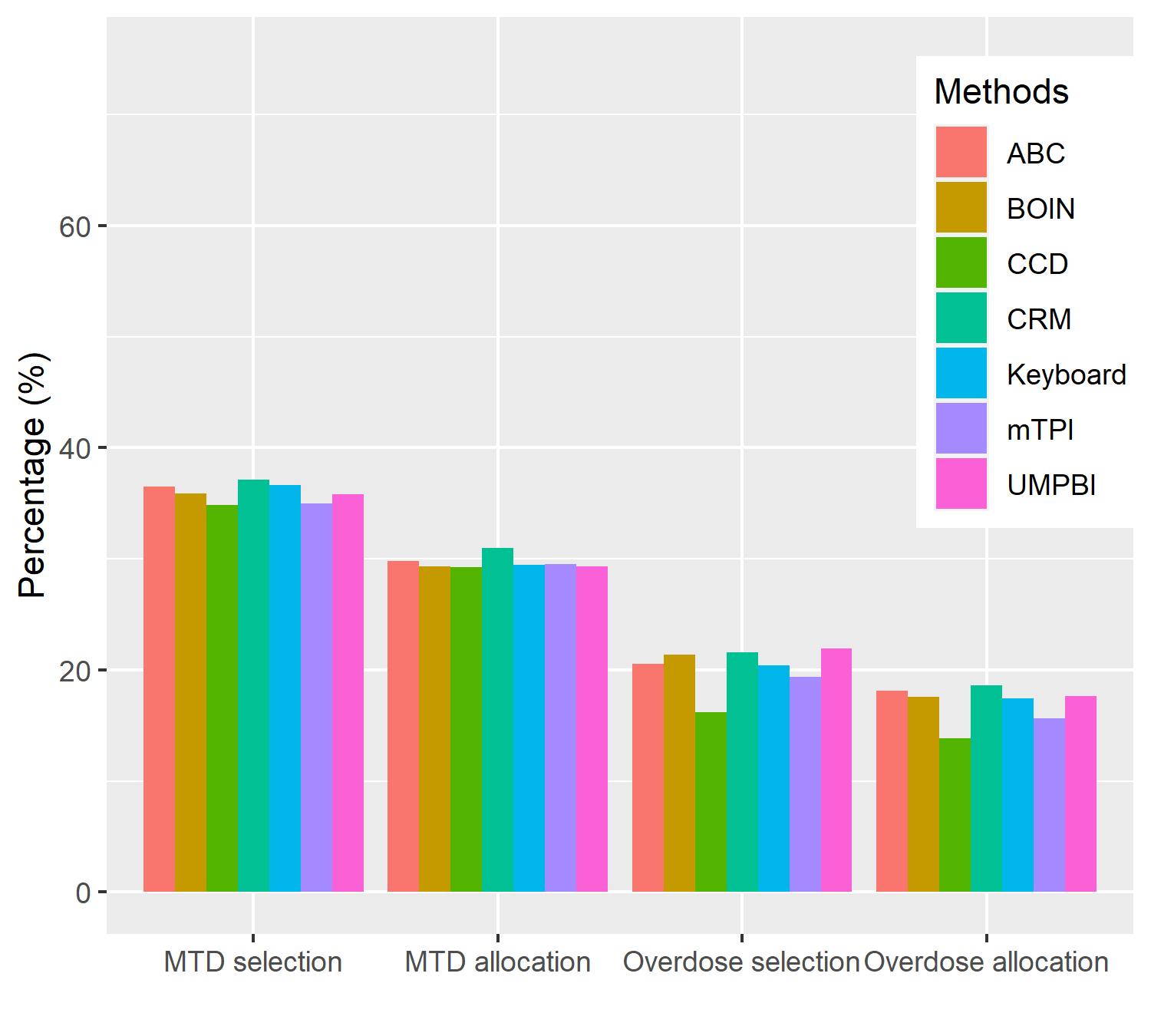}  
		\caption{$\Delta=0.05$}
	\end{subfigure}
	\begin{subfigure}{0.5\textwidth}
		\centering
		% include second image
		\includegraphics[width=.9\linewidth]{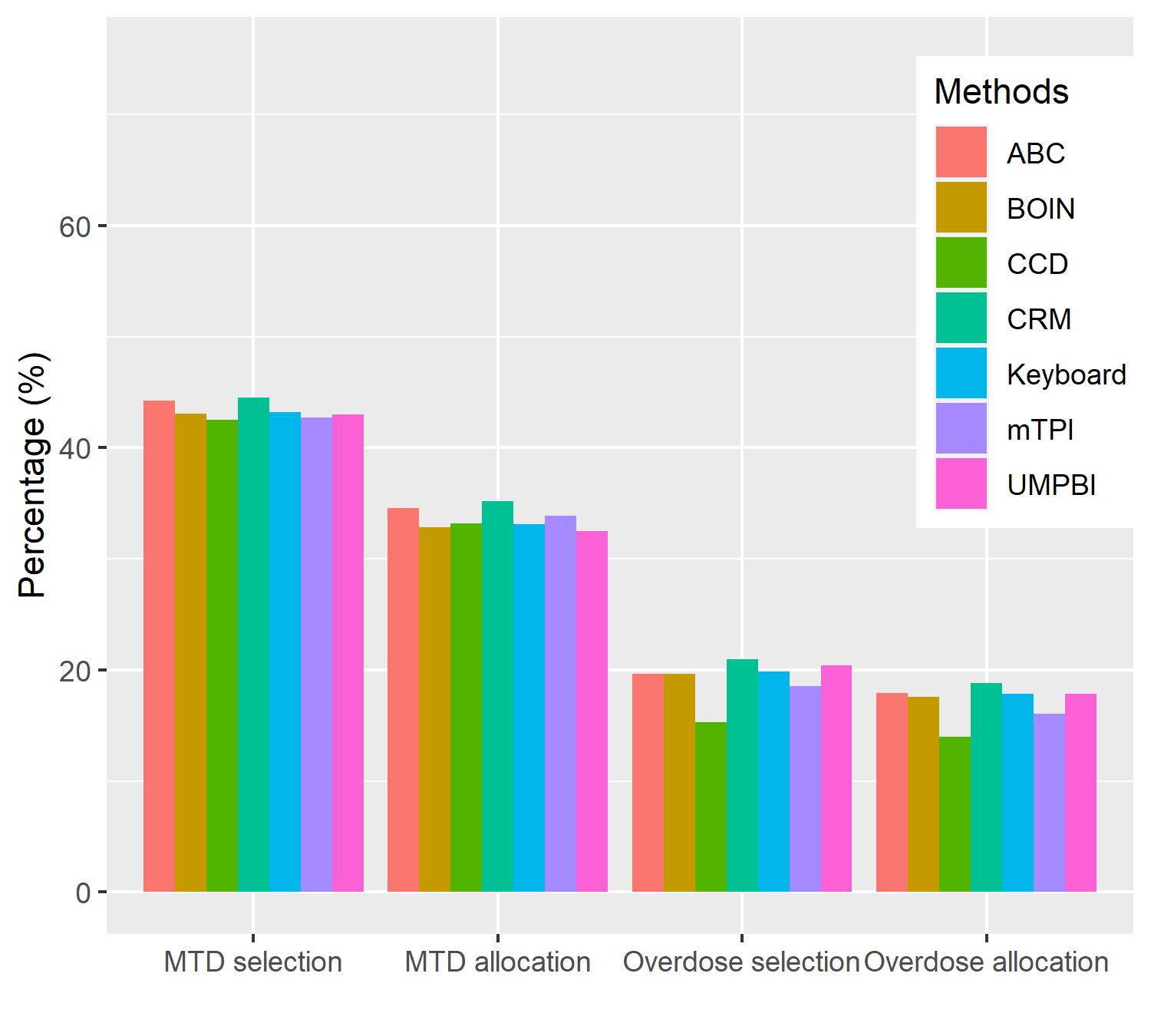}
		\subcaption{$\Delta=0.07$}
	\end{subfigure}
	
	\begin{subfigure}[]{0.5\textwidth}
		\centering
		% include first image
		\includegraphics[width=.9\linewidth]{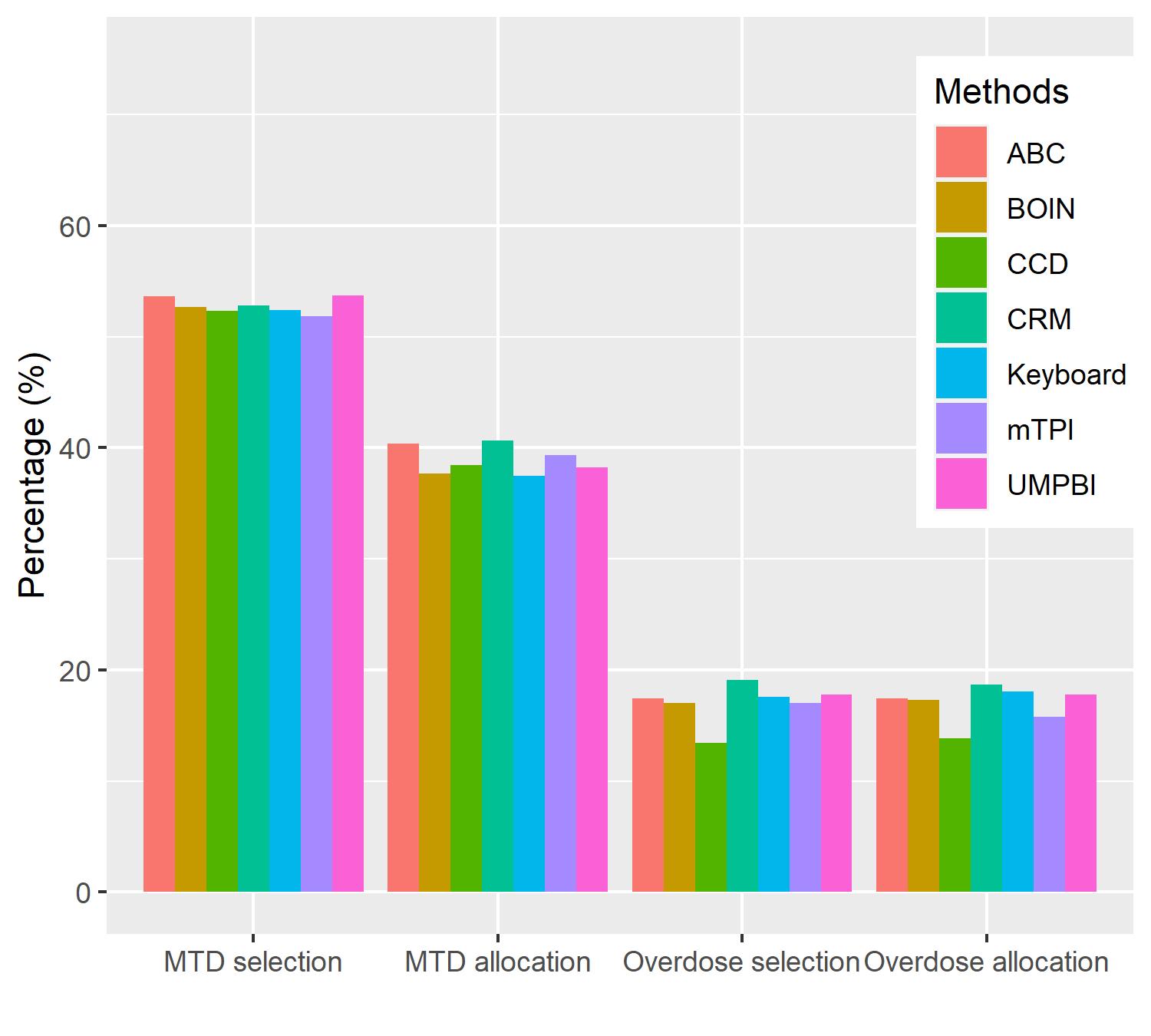}  
		\caption{$\Delta=0.10$}
	\end{subfigure}
	\begin{subfigure}{0.5\textwidth}
		\centering
		% include second image
		\includegraphics[width=.9\linewidth]{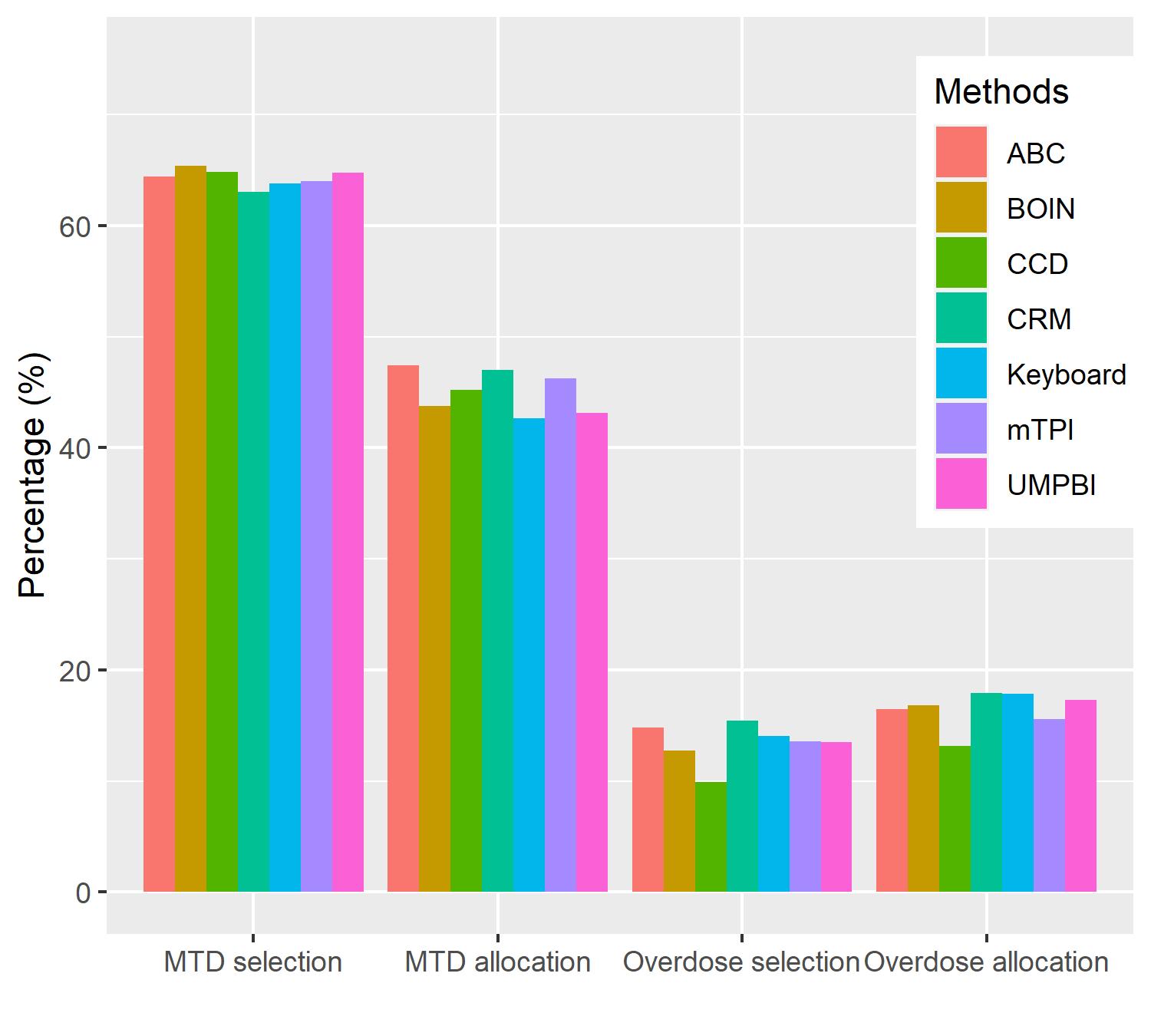}
		\subcaption{$\Delta=0.15$}
	\end{subfigure}
	%\newline
	
	\caption{Simulation results 
	%for the dose-finding trials 
	with sample size $30$ based on 5000 randomly generated
		dose–toxicity scenarios under the average probability difference of $\Delta=0.05$, 0.07, 0.10 and 0.15
		around the target toxicity probability $\phi= 0.30$, respectively.}
	\label{fig:cpr30}
\end{figure}

As shown in Figure~\ref{fig:cpr30}, 
all the methods yield better results when $\Delta$ increases.
The efficiency of using all the available data under the ABC and CRM designs 
is reflected by the metric MTD allocation, 
where both designs yield higher MTD allocation percentages compared with the other interval designs.
The gap becomes more significant when $\Delta$ increases. 
In terms of the MTD selection percentage,
both ABC and CRM designs are superior to the interval designs 
when $\Delta$ is small. 
When $\Delta$ becomes large, the gap diminishes.
It is possibly because when $\Delta$ is large, 
the information from data is adequate to identify the MTD well,
thus using all the available data does not boost the performance significantly.
It is also worth noting that 
when $\Delta=0.15$, 
the CRM design has a bit lower MTD selection percentage compared with the counterparts
while the ABC design is robust across different $\Delta$'s.
This reveals the advantage of imposing no dose--toxicity model assumption for the ABC design.

With regard to the two safety measurements, 
overall the CRM design has the highest overdose selection and allocation percentages
among the seven designs.
The CCD design performs the best for the safety metrics while 
it leads to worse results for the two main measurements.
For the other five designs, 
they are comparable in terms of the safety metrics.
In summary, the comparisons with the six well-known methods demonstrate the robustness and efficiency of the ABC design.

\subsection{Evaluation under Fixed Scenarios}\label{subsec:fix}
To gain more insight 
%obtain a further investigation 
into the ABC design, 
we evaluate its performance under five representative fixed scenarios.
For an objective comparison and avoiding cherry picking, we adopt the fixed scenarios in Table~1 of 
Cheung and Chappell \citep{cheung2000sequential}.
The number of dose levels is $K=6$ and the target toxicity rate is $\phi=0.2$.
The comparisons are still made with the six phase I designs under the similar settings as introduced in Section~A.1 of the Appendix.
We choose the total sample size as $36$ 
and the cohort size is fixed at $3$.
We use the same early stopping rules as under the random scenarios.
The detailed information of the fixed scenarios can be found in Table~\ref{simu:fix36}.
For each scenario, we replicate 5000 simulations. 

The results are presented in Table~\ref{simu:fix36}.
For scenario 1, the CRM has the best performance, while
the other designs yield comparable results.
In scenario 2, all the dose levels are overly toxic, and the ABC design and all interval designs show satisfactory and comparable results by early stopping the trials.
Only the CRM method displays a slightly lower non-selection percentage ($53.4\%$) compared with others.
The ABC design leads to the best performance under both scenarios 3 and 4
and the gap of MTD selection percentage is around $10\%$ under scenario 3.
Under scenario 6 where the last dose level is the MTD,
the mTPI design is significantly better than others,
but it is worth noting that the mTPI design tends to select over-toxic dose level as the MTD 
in our simulation studies
(see scenarios 1, 3 and 4).
Except for the mTPI design, 
the ABC design has slightly higher over-dose selection percentages under scenarios 1, 3 and 4, while
the gap is marginal.
Overall, the ABC design yields satisfactory performances for all the five fixed scenarios.

\section{Real Trial Application}\label{sec:realdata}

As an illustration, we 
apply the ABC design to the aforementioned phase I trial on the MEK inhibitor selumetinib in children with progressive LGG. The DLT outcomes were defined as any grade 4 toxicity (except lymphopenia), grade 3 neutropenia with fever, grade 3 thrombocytopenia with bleeding,  any grade 3 or 4 toxicity possibly related to selumetinib, or any grade 2 toxicity persisting $\ge 7$ days that was medically significant or intolerable enough to interrupt or reduce the dose. 
Originally, the trial evaluated $37$ patients to estimate the MTD with the target toxicity rate $\phi=0.25$.
Patients were grouped in a cohort size of $3$ to be assigned to one of the three dose levels of the MEK inhibitor selumetinib $\{25, 33, 43\}$ mg/m$^2$/dose bis in die via the CRM design based on the two-parameter logistic model. 
The observed data from the original clinical trial were 
\bse
\{y_1, y_2, y_3\} &=& \{3, 4, 2\} \\
\{m_1, m_2, m_3\} &=& \{24, 10, 3\}.
\ese
The MTD selected using the CRM method is dose level 1\citep{banerjee2017phase}. 
%However, with only three dose levels under consideration, it may not be reliable to use the CRM model to regress on three doses. 
Based on the observed data in the trial, the estimated DLT rates were $\{0.125, 0.400, 0.667\}$, respectively.

% \begin{figure}[!t]
% 	\centering
% 	\includegraphics[width=1\columnwidth]{real_data.png}
% 	\caption{
% 	Patient allocations and outcomes by the ABC design for the real trial illustration using a representative trial among the $5000$ repetitions. The ABC recommended MTD is dose level 1.
% 	}
% 	\label{fig:realdata}
% \end{figure}

We reran this trial based on the estimated DLT rates by the ABC design
with $\delta=0.1$ and $h=0.01$.
There were 13 cohorts in total where the last cohort contained only one subject.
For comparison, we also include the CRM design using the two-parameter logistic model,
for which the detailed setting is given in Section~A.1 of the Appendix.
The entire procedure is repeated for $5000$ times 
and the results are presented in Table~\ref{tab:real}.
It is clear that the ABC design yields comparable performances to the CRM design under this trial example.
Nevertheless, the ABC design is model-free, 
and thus it is more robust and easier to use in practice.

% \begin{table}
% 	\caption{
% 	 The percentages of MTD selection (the numbers of patients treated at each dose)
% under the ABC and CRM designs with 5000 replications. The target toxicity probability is 0.25 in the real trial application with 37 patients.}
% 	\label{tab:real}
% 	\centering
% 	\renewcommand{\arraystretch}{0.8}
% 	\begin{tabular}{lccccc}
% 		\toprule
%  		Design &\multicolumn{3}{c}{Dose Level }& DLT&  None \\
% 		&1&2&3&(\%) & (\%)\\ \midrule
% 		 &\textbf{0.12}&0.40&0.67&\\
% ABC & 55.9 (19.3) & 43.4 (16.6) & 0.2 (0.9) & 26.2 &  0.6 \\
% CRM & 56.4 (18.8) & 42.7 (16.5) & 0.1 (1.5) & 26.9 &  0.8\\
% 
% \bottomrule
% 	\end{tabular}
% \end{table}

\begin{figure}[!ht]
	\centering
	\includegraphics[width=1\columnwidth]{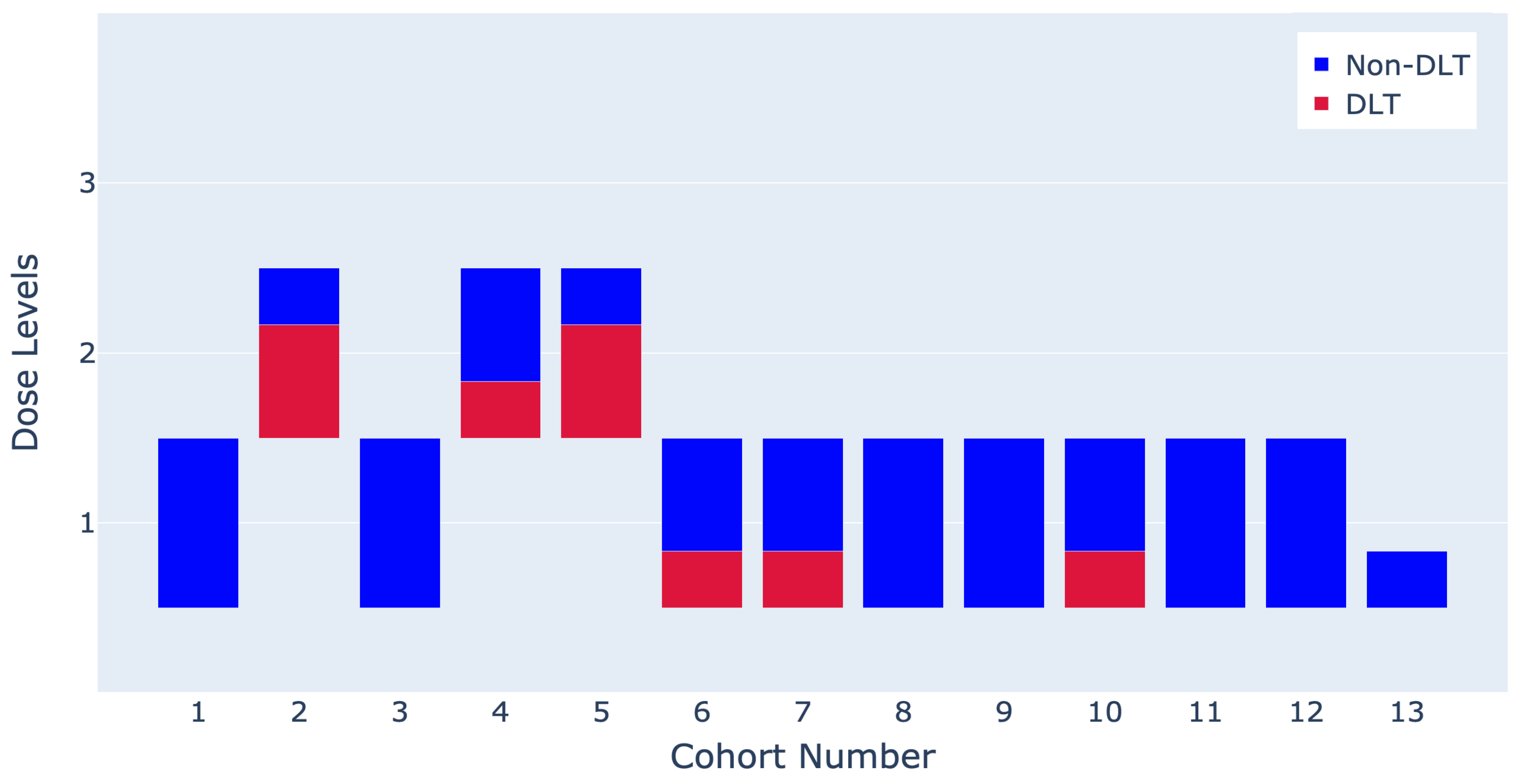}
	\caption{
	Patient allocations and outcomes by the ABC design for the real trial illustration using a representative trial among the $5000$ repetitions. The ABC recommended MTD is dose level 1.
	}
	\label{fig:realdata}
\end{figure}

To better demonstrate the property of the ABC design, 
we present the detailed procedure of one specific trial selected from the $5000$ repetitions.
The patient allocations and outcomes are shown in Figure~\ref{fig:realdata}.
The trial started treating the first cohort of patients at the lowest dose level. 
For the first cohort, 
there was no DLT observed, 
which yielded the estimated DLT rates as 
$\{\hat{p}_{1, k}\}_{k=1}^3= (0.08, 0.22, 0.40)$.
Thus, the trial escalated to dose level 2 for the second cohort,
where we observed two DLTs among three patients, resulting in the estimated DLT rates $\{\hat{p}_{2, k}\}_{k=1}^3=(0.18, 0.37, 0.45)$.
Consequently, the next cohort was assigned back to dose level 1, where no patient experienced the DLT outcome, leading to
$\{\hat{p}_{3, k}\}_{k=1}^K= (0.12, 0.33, 0.44)$.
The trial then escalated to dose level 2 and
one out of the three patients experienced the DLT for this cohort, which led to $\{\hat{p}_{4, k}\}_{k=1}^K= (0.11, 0.33, 0.44)$.
Consequently, the trial stayed at dose level 2 again.
There were two DLT outcomes among the three patients 
and the trial de-escalated to dose level 1.
All the remaining cohorts were assigned to dose level 1.
Finally, the observed data were 
\bse
\{y_1, y_2, y_3\} &=& \{3, 5, 0\} \\
\{m_1, m_2, m_3\} &=& \{28, 9, 0\}.
\ese
Upon the completion of the trial,
we selected dose level 1 (i.e., the dose of 25 mg/m$^2$/dose bis in die) as the
MTD,
which is consistent with the original selection by the CRM design \citep{banerjee2017phase}.

\section{Conclusion}\label{sec:con}
We have proposed a new phase I design for identifying the MTD with the approximate Bayesian computation method.
The ABC design possesses the merits of both model-free and model-based designs simultaneously.
Because it is model-free, there is no need to specify any assumption on the dose--toxicity curve, which avoids the risk of model misspecification.
%At the same time, 
Similar to the model-based methods, the ABC design is also efficient by aggregating all the available information when deciding for dose movement for each new cohort.

The extensive simulation studies indicate that the ABC design is efficient and robust under different trial settings.
Compared with other phase I designs, 
the ABC design shows advantages in terms of the MTD selection and patient allocation under the random scenarios.
There are two tuning parameters $\delta$ and $h$ in the ABC design.
The neighborhood parameter $\delta$ has minor effect on the design performance, as shown by the comprehensive simulation studies and ANOVA in Sensitivity Analysis Section,
while 
the bandwidth parameter $h$ can be easily selected as $h=0.01$ 
to achieve satisfactory performance.
Therefore, in practice, the ABC design is easy to use, as there is no need to carry out the extensive parameter calibration prior to the trial conduct. 
The ABC design can be broadly used in phase I clinical trials due to its robustness property and ease of implementation.
%it takes more advantage of the efficient property in selecting MTD when the total sample size is not too small.
Under some cases in the simulation studies,
our design shows a slightly higher but acceptable overdose selection percentage compared with existing phase I designs as a sacrifice for higher MTD selection and MTD allocation percentages.
This is a trade-off often encountered in dose finding: by exploring more doses, it would help to pin down the MTD more accurately, while more patients might be put at the risk of exploring untried (higher) dose levels.
%design in our simulation studies.
The ABC design may not be applicable to 
phase I trials with the possibility of inserting some intermediate dose level,
because a dose--toxicity model is typically needed for such dose insertion. Nevertheless, one possibility along this direction is to incorporate a working model to pin down the MTD more precisely through interpolation and dose insertion.
%However, the main advantage of the ABC design is that it is model-free.

The ABC design can be easily extended to other more complicated phase I trials.
To account for the late-onset outcome, the ABC can be combined with the fractional design \citep{yin2013fractional} in a straightforward way.
The only modification is to tune the bandwidth parameter $h$ under the late-onset context.
To accommodate the efficacy outcomes, 
we can 
introduce the admissible set $\mathcal{A}$ and 
use a beta--binomial model to estimate the efficacy rate
and thus deliver decisions through a trade-off between toxicity and efficacy.
%take the efficacy outcomes into consideration.
The ABC design can also be extended to the dose combination trials with an adaptation on generating prior samples, which warrants further investigation.

We have developed an application (\href{https://github.com/JINhuaqing/ABC}{https://github.com/JINhuaqing/ABC}) for dose finding based on our ABC design, where users can set various customized configurations for their trials and obtain visualization results of the ABC design. 
The R scripts for simulation studies as well as the real data application
are available at \href{https://github.com/JINhuaqing/ABC-simu}{https://github.com/JINhuaqing/ABC-simu}.
The implementation of the ABC design is simple and fast
due to the fact that the prior samples can be generated beforehand.

	\bibliographystyle{apa.bst}
	\bibliography{ref}%

\newpage

\clearpage
\begin{table}[!ht]
	\caption{
	The simulation factors that may affect the dose-finding performance of phase I
trial design and the results of ANOVA in terms of the percentage of MTD selection. The ANOVA
also includes all the pairwise interactions between the five simulation factors.
}
	\label{anova}
	\centering
    \resizebox{\linewidth}{!}{
	\begin{tabular}{lcrrr}
		\toprule
		Factors &Levels of factor & DF & SS &  MSE\\
		\midrule
		Average probability difference around $\phi$ ($\Delta$) &\{0.05, 0.07, 0.10, 0.15\} &3& 65.35 & 21.78 \\
		Bandwidth parameter $h$ & \{0.1, 0.05, 0.01, 0.005\}&4& 4.32 & 1.44\\
		Sample size&$\{18, 24, \dots,60\}$ & 7& 5.58 & 0.80 \\
		Number of dose levels $K$ &\{3, 5,  7\} & 2 & 0.81 & 0.40\\
		Target toxicity probability $\phi$ &\{0.25, 0.30, 0.33\}&2& 0.34 & 0.17\\
		Neighborhood parameter $\delta$ & \{0.0, 0.05, 0.10, 0.15, 0.20, random\}&5& 0.54 & 0.11\\
		\midrule
		Total variance &  & 6911 &  83.39 &  \\
		\bottomrule
		\multicolumn{5}{l}{DF: degree of freedom; SS: sum of squares; MSE: mean squared error (MSE$=$SS/DF)}
	\end{tabular}
	}
\end{table}

\begin{table}[!ht]
	\caption{
	 The percentages of MTD selection (the numbers of patients treated at each dose)
under the ABC design in comparison with the BOIN, CCD, CRM, Keyboard, mTPI and UMPBI designs under six fixed scenarios with the target toxicity probability 0.20 in boldface when sample size is $36$.
None represents the percentage of trials of non-selection  due to early stopping.
}
   \label{simu:fix36}
	\centering
	\renewcommand{\arraystretch}{0.8}
	\resizebox{\textwidth}{!}{
	\begin{tabular}{lcccccccc}
		\toprule
 		&\multicolumn{6}{c}{Dose Level }& DLT& None\\
		Design &1&2&3&4&5&6&(\%) & (\%)\\
        \midrule
Scenario 1&0.05&0.10&\textbf{0.20}&0.30&0.50&0.70& \\
ABC      & 1.1 (4.2) & 21.3 (9.0)  & 49.7 (12.8) & 25.1 (7.9) & 2.0 (1.7) & 0   (0.1) & 19.4 & 0.8 \\ 
BOIN     & 0.8 (6.1) & 25.0 (11.1) & 52.0 (11.8) & 20.4 (5.5) & 1.5 (1.2) & 0.1 (0.1) & 17.1 & 0.2 \\
CCD      & 3.2 (6.8) & 30.1 (11.9) & 49.0 (11.8) & 16.1 (4.5) & 1.2 (0.9) & 0   (0.1) & 16.0 & 0.3 \\
CRM      & 0.8 (5.2) & 22.2 (9.9)  & 56.5 (13.2) & 19.7 (6.5) & 0.8 (1.1) & 0   (0.1) & 18.0 & 0.0 \\
Keyboard & 1.4 (6.2) & 25.0 (11.2) & 52.1 (11.7) & 19.8 (5.5) & 1.5 (1.3) & 0   (0.1) & 17.0 & 0.2 \\
mTPI     & 1.0 (4.8) & 17.3 (8.3)  & 46.0 (12.5) & 33.1 (8.3) & 2.3 (1.9) & 0   (0.1) & 19.8 & 0.3 \\
UMPBI    & 0.7 (5.9) & 24.7 (10.8) & 50.5 (12.0) & 22.1 (5.8) & 1.6 (1.3) & 0.1 (0.1) & 17.4 & 0.2 \\
        \midrule
Scenario 2&0.30&0.40&0.52&0.61&0.76&0.87& \\
ABC      & 39.1 (16.9) & 3.7 (4.5) & 0.1 (0.8) & 0 (0.1)  & 0 (0) & 0 (0) & 32.7 &  57.2 \\ 
BOIN     & 37.4 (18.6) & 3.5 (3.6) & 0.1 (0.5) & 0 (0)    & 0 (0) & 0 (0) & 32.3 &  58.9 \\
CCD      & 39.6 (19.9) & 2.5 (2.8) & 0.1 (0.4) & 0 (0)    & 0 (0) & 0 (0) & 31.6 &  57.9 \\
CRM      & 44.4 (22.5) & 2.1 (3.2) & 0.1 (0.6) & 0 (0.1)  & 0 (0) & 0 (0) & 31.7 &  53.4 \\
Keyboard & 38.9 (18.8) & 3.4 (3.7) & 0.1 (0.5) & 0 (0)    & 0 (0) & 0 (0) & 32.0 &  57.6 \\
mTPI     & 36.6 (17.0) & 7.2 (5.5) & 0.3 (0.8) & 0 (0.1)  & 0 (0) & 0 (0) & 33.3 &  55.9 \\
UMPBI    & 39.8 (18.9) & 3.5 (3.8) & 0.1 (0.6) & 0 (0)    & 0 (0) & 0 (0) & 32.1 &  56.6 \\
        \midrule
Scenario 3&0.05&0.06&0.08&0.11&\textbf{0.19}&0.34& \\
ABC      & 0.3 (3.8) & 1.4 (4.4) &  4.6 (5.2) & 23.3 (8.1) & 54.0 (11.1)& 15.6 (3.3) & 14.0&  0.8\\ 
BOIN     & 0.3 (5.2) & 3.3 (5.8) & 10.2 (6.6) & 27.5 (7.6) & 43.0 (7.1) & 15.3 (3.6) & 12.8&  0.3\\
CCD      & 1.8 (5.9) & 7.8 (6.5) & 14.6 (6.9) & 27.6 (7.4) & 37.4 (6.4) & 10.4 (2.8) & 11.7&  0.3\\
CRM      & 0.2 (4.5) & 3.1 (5.2) & 11.6 (6.6) & 29.2 (8.2) & 43.2 (8.0) & 12.7 (3.7) & 13.1&  0.1\\
Keyboard & 0.5 (5.2) & 3.5 (5.7) & 10.6 (6.7) & 27.0 (7.6) & 43.2 (7.1) & 15.0 (3.6) & 12.8&  0.3\\
mTPI     & 0.5 (4.6) & 2.9 (4.8) &  7.2 (5.9) & 21.8 (6.8) & 44.9 (8.1) & 22.5 (5.7) & 14.6&  0.2\\
UMPBI    & 0.3 (5.0) & 3.0 (5.6) &  9.4 (6.5) & 28.5 (7.7) & 42.1 (7.3) & 16.6 (3.8) & 12.8&  0.2\\
        \midrule
Scenario 4&0.06&0.08&0.12&\textbf{0.18}&0.40&0.71& \\
ABC      & 0.7 (4.2) &  5.1 (5.6) & 21.9 (8.2) & 57.5 (12.4) & 13.5 (5.1) & 0.3 (0.2) & 17.2 & 1.0\\ 
BOIN     & 0.9 (5.9) &  8.8 (7.5) & 28.0 (9.1) & 49.6 (8.9)  & 11.9 (4.0) & 0.2 (0.5) & 15.7 & 0.5\\
CCD      & 3.8 (7.1) & 15.1 (8.4) & 28.9 (9.0) & 43.3 (8.2)  &  8.3 (2.9) & 0.1 (0.3) & 14.1 & 0.4\\
CRM      & 0.7 (5.1) &  7.5 (6.5) & 30.2 (9.6) & 51.6 (10.8) &  9.9 (3.6) & 0.1 (0.3) & 15.6 & 0.1\\
Keyboard & 1.3 (5.9) &  8.6 (7.4) & 27.1 (9.0) & 50.0 (8.9)  & 12.2 (4.1) & 0.3 (0.5) & 15.6 & 0.5\\
mTPI     & 1.3 (5.0) &  6.6 (6.0) & 18.4 (7.9) & 56.9 (10.5) & 16.2 (5.9) & 0.1 (0.6) & 17.8 & 0.5\\
UMPBI    & 0.8 (5.8) &  8.2 (7.4) & 26.1 (8.9) & 52.2 (9.2)  & 12.1 (4.1) & 0.2 (0.5) & 15.8 & 0.4\\

        \midrule
Scenario 5&0.00&0.00&0.03&0.05&0.11&\textbf{0.22}& \\
ABC      & 0 (3.0)  & 0 (3.0) & 0.1 (3.4) & 2.5 (4.6) & 37.6 (11.2)& 59.8 (10.8) & 10.9 & 0 \\ 
BOIN     & 0 (3.0)  & 0 (3.3) & 0.3 (4.2) & 5.4 (6.0) & 36.5 (9.1) & 57.8 (10.4) & 10.3 & 0 \\ 
CCD      & 0 (3.0)  & 0 (3.3) & 1.5 (4.7) & 8.6 (6.4) & 40.1 (9.2) & 49.9 (9.4)  & 9.8  & 0 \\ 
CRM      & 0 (3.0)  & 0 (3.0) &   0 (3.4) & 2.9 (4.8) & 34.4 (9.1) & 62.6 (12.6) & 11.4 & 0 \\ 
Keyboard & 0 (3.0)  & 0 (3.3) & 0.5 (4.3) & 5.8 (6.0) & 35.0 (8.9) & 58.7 (10.5) & 10.3 & 0 \\ 
mTPI     & 0 (3.0)  & 0 (3.0) & 0.4 (3.8) & 3.6 (4.7) & 27.3 (7.7) & 68.8 (13.8) & 11.9 & 0 \\ 
UMPBI    & 0 (3.0)  & 0 (3.3) & 0.5 (4.3) & 5.7 (5.9) & 36.3 (9.0) & 57.5 (10.5) & 10.4 & 0 \\ 

        \bottomrule
	\end{tabular}
	}
\end{table}

\begin{table}[!ht]
	\caption{
	 The percentages of MTD selection (the numbers of patients treated at each dose)
under the ABC and CRM designs with 5000 replications. The target toxicity probability is 0.25 in the real trial application with 37 patients.}
	\label{tab:real}
	\centering
	\renewcommand{\arraystretch}{0.8}
	\begin{tabular}{lccccc}
		\toprule
 		Design &\multicolumn{3}{c}{Dose Level }& DLT&  None \\
		&1&2&3&(\%) & (\%)\\ \midrule
		 &\textbf{0.12}&0.40&0.67&\\
ABC & 55.9 (19.3) & 43.4 (16.6) & 0.2 (0.9) & 26.2 &  0.6 \\
CRM & 56.4 (18.8) & 42.7 (16.5) & 0.1 (1.5) & 26.9 &  0.8\\

\bottomrule
	\end{tabular}
\end{table}

\end{document}

\def\eqx"#1"{{\label{#1}}}
\def\eqn"#1"{{\ref{#1}}}

\makeatletter % make @ act like a letter
\@addtoreset{equation}{section}
\makeatother  % make @ act like a non-letter

\def\yincomment#1{\vskip 2mm\boxit{\vskip 2mm{\color{red}\bf#1} {\color{blue}\bf --Yin\vskip 2mm}}\vskip 2mm}
\def\squarebox#1{\hbox to #1{\hfill\vbox to #1{\vfill}}}
\def\boxit#1{\vbox{\hrule\hbox{\vrule\kern6pt
          \vbox{\kern6pt#1\kern6pt}\kern6pt\vrule}\hrule}}

\newcommand{\blue}[1]{\textcolor{blue}{{#1}}}
\newcommand{\red}[1]{\textcolor{red}{{#1}}}
\def\theequation{\thesection.\arabic{equation}}
\newcommand{\ds}{\displaystyle}

\newcommand{\bJ}{\mbox{\bf J}}
\newcommand{\bF}{\mbox{\bf F}}
\newcommand{\bM}{\mbox{\bf M}}
\newcommand{\bR}{\mbox{\bf R}}
\newcommand{\bZ}{\mbox{\bf Z}}
\newcommand{\bX}{\mbox{\bf X}}
\newcommand{\bx}{\mbox{\bf x}}
\newcommand{\bQ}{\mbox{\bf Q}}
\newcommand{\bH}{\mbox{\bf H}}
\newcommand{\bh}{\mbox{\bf h}}
\newcommand{\bz}{\mbox{\bf z}}
\newcommand{\ba}{\mbox{\bf a}}
\newcommand{\bG}{\mbox{\bf G}}
\newcommand{\bB}{\mbox{\bf B}}
\newcommand{\bb}{\mbox{\bf b}}
\newcommand{\bA}{\mbox{\bf A}}
\newcommand{\bC}{\mbox{\bf C}}
\newcommand{\bI}{\mbox{\bf I}}
\newcommand{\bD}{\mbox{\bf D}}
\newcommand{\bU}{\mbox{\bf U}}
\newcommand{\bc}{\mbox{\bf c}}
\newcommand{\bd}{\mbox{\bf d}}
\newcommand{\bs}{\mbox{\bf s}}
\newcommand{\bS}{\mbox{\bf S}}
\newcommand{\bV}{\mbox{\bf V}}
\newcommand{\bv}{\mbox{\bf v}}
\newcommand{\bW}{\mbox{\bf W}}
\newcommand{\bw}{\mbox{\bf w}}
\newcommand{\bg}{\mbox{\bf g}}
\newcommand{\bu}{\mbox{\bf u}}

\newcommand{\bcU}{\boldsymbol{\cal U}}
\newcommand{\bbeta}{\boldsymbol{\beta}}
\newcommand{\bdelta}{\boldsymbol{\delta}}
\newcommand{\bDelta}{\boldsymbol{\Delta}}
\newcommand{\boldeta}{\boldsymbol{\eta}}
\newcommand{\bxi}{\boldsymbol{\xi}}
\newcommand{\bGamma}{\boldsymbol{\Gamma}}
\newcommand{\bSigma}{\boldsymbol{\Sigma}}
\newcommand{\balpha}{\boldsymbol{\alpha}}
\newcommand{\bOmega}{\boldsymbol{\Omega}}
\newcommand{\btheta}{\boldsymbol{\theta}}
\newcommand{\bmu}{\boldsymbol{\mu}}
\newcommand{\bnu}{\boldsymbol{\nu}}
\newcommand{\bgamma}{\boldsymbol{\gamma}}

\newcommand{\bse}{\begin{eqnarray*}}
\newcommand{\ese}{\end{eqnarray*}}
\newcommand{\be}{\begin{eqnarray}}
\newcommand{\ee}{\end{eqnarray}}
\newcommand{\bsq}{\begin{equation*}}
\newcommand{\esq}{\end{equation*}}
\newcommand{\bq}{\begin{equation}}
\newcommand{\eq}{\end{equation}}
\newcommand{\var}{\mbox{var}}
\newcommand{\trace}{\hbox{trace}}
\newcommand{\wh}{\widehat}
\newcommand{\wt}{\widetilde}
\newcommand{\eff}{_{\rm eff}}
\newcommand{\sub}{{\rm sub}}
\newcommand{\cat}{{\rm cat}}
\newcommand{\eLL}{\mathcal L}
\newcommand{\n}{\nonumber}
\newcommand{\bias}{\mbox{bias}}
\newcommand{\vecl}{\mbox{vecl}}
\newcommand{\AIC}{\mbox{AIC}}
\newcommand{\BIC}{\mbox{BIC}}
\newcommand{\MSE}{\mbox{MSE}}
\newcommand{\rank}{\mbox{rank}}
\newcommand{\cov}{\mbox{cov}}
\newcommand{\corr}{\mbox{corr}}
\newcommand{\argmin}{\mbox{argmin}}
\newcommand{\argmax}{\mbox{argmax}}
\newcommand{\diag}{\mbox{diag}}
\newcommand{\trans}{^{\rm \top}}
\newcommand{\bTheta}{\boldsymbol\Theta}
\newcommand{\bta}{\boldsymbol\eta}
\newcommand{\bphi}{\boldsymbol\phi}
\newcommand{\btau}{\boldsymbol\tau}
\newcommand{\boeta}{\boldsymbol\eta}
\newcommand{\bpsi}{\boldsymbol\psi}
\newcommand{\0}{{\bf 0}}
\newcommand{\A}{{\bf A}}
\newcommand{\U}{{\bf U}}
\newcommand{\V}{{\bf V}}
\newcommand{\e}{{\bf e}}
\newcommand{\R}{{\bf R}}
\newcommand{\G}{{\bf G}}
\newcommand{\bO}{{\bf O}}
\newcommand{\B}{{\bf B}}
\newcommand{\D}{{\bf D}}
\newcommand{\K}{{\bf K}}
\newcommand{\g}{{\bf g}}
\newcommand{\f}{{\bf f}}
\newcommand{\h}{{\bf h}}
\newcommand{\I}{{\bf I}}
\newcommand{\M}{\mbox{ $\mathcal{M}$}}
\newcommand{\BB}{\mbox{ $\mathcal{B}$}}
\newcommand{\N}{\mbox{ $\mathcal{N}$}}
\newcommand{\T}{{\bf T}}
\newcommand{\bP}{{\bf P}}
\newcommand{\s}{{\bf s}}
\newcommand{\m}{{\bf m}}
\newcommand{\W}{{\bf W}}
\newcommand{\w}{{\bf w}}
\newcommand{\X}{{\bf X}}
\newcommand{\x}{{\bf x}}
\newcommand{\tx}{{\widetilde \x}}
\newcommand{\Y}{{\bf Y}}
\newcommand{\C}{{\bf C}}
\newcommand{\tY}{{\widetilde Y}}
\newcommand{\y}{{\bf y}}
\newcommand{\Z}{{\bf Z}}
\newcommand{\z}{{\bf z}}
\newcommand{\Ybar}{{\overline{Y}}}
\newcommand{\Xbar}{{\overline{\X}}}
\newcommand{\xbar}{{\overline{\x}}}
\newcommand{\wbar}{{\overline{\W}}}
\newcommand{\bSig}{{\bf \Sigma}}
\newcommand{\bLam}{{\bf \Lambda}}
\def\th{^{th}}
\def\S{{\bf S}}
\def\L{{\bf L}}
\def\u{{\bf u}}
\def\v{{\bf v}}
\def\T{{\bf T}}
\def\bO{{\bf O}}
\def\I{{\bf I}}
\def\K{{\bf K}}
\def\t{{\bf t}}
\def\b{{\bf b}}
\def\r{{\bf r}}
\def\V{{\bf V}}
\def\c{{\bf c}}
\def\a{{\bf a}}
\def\vec{\mbox{vec}}

\newcommand{\cvec}[1]{{\mathbf #1}}
\newcommand{\rvec}[1]{\vec{\mathbf #1}}
\newcommand{\minor}{{\rm minor}}
\newcommand{\spn}{{\rm Span}}
\newcommand{\range}{{\rm range}}
\newcommand{\mdiv}{{\rm div}}
\newcommand{\proj}{{\rm proj}}
\newcommand{\RR}{\mathbb{R}}
\newcommand{\NN}{\mathbb{N}}
\newcommand{\QQ}{\mathbb{Q}}
\newcommand{\ZZ}{\mathbb{Z}}
\newcommand{\EE}{\mathbb{E}}
\newcommand{\<}{\langle}
\renewcommand{\>}{\rangle}
\renewcommand{\emptyset}{\varnothing}
\newcommand{\attn}[1]{\textbf{#1}}
\newcommand{\bproof}{\bigskip {\bf Proof. }}
\newcommand{\eproof}{\hfill\qedsymbol}
\newcommand{\Disp}{\displaystyle}
\newcommand{\qe}{\hfill\(\bigtriangledown\)}
\newcommand*{\dif}{\mathop{}\!\mathrm{d}}

\newtheorem{thm}{Theorem}[section]
\newtheorem{lem}{Lemma}[section]
\newtheorem{rem}{Remark}[section]
\newtheorem{cor}{Corollary}[section]
\newcolumntype{L}[1]{>{\raggedright\let\newline\\\arraybackslash\hspace{0pt}}m{#1}}
\newcolumntype{C}[1]{>{\centering\let\newline\\\arraybackslash\hspace{0pt}}m{#1}}
\newcolumntype{R}[1]{>{\raggedleft\let\newline\\\arraybackslash\hspace{0pt}}m{#1}}

\newcommand{\tabincell}[2]{\begin{tabular}{@{}#1@{}}#2\end{tabular}}

\newtheorem{theorem}{Theorem}
\newtheorem{definition}{Definition}

\newcommand{\dT}{\top}

\newcommand{\algorithmicobs}{\textbf{Observations:}}
\newcommand{\algorithmicprior}{\textbf{Prior:}}
\newcommand{\PRIOR}{\item[\algorithmicprior]}
\newcommand{\OBS}{\item[\algorithmicobs]}

\newcommand{\algorithmicoutput}{\textbf{Output:}}
\newcommand{\OUTPUT}{\item[\algorithmicoutput]}

\newcommand{\whtheta}{\widehat{\theta}}
\newcommand{\htheta}{\hat{\theta}}
\newcommand{\UIP}{\mathrm{UIP}}
\newcommand{\pmean}{\mu}
\newcommand{\pvar}{\eta^2}
\newcommand{\UI}{\mathrm{I}_{\mathrm{U}}}
\newcommand{\MLE}{\mathrm{MLE}}
\newcommand{\init}{\mathrm{ini}}
\newcommand{\med}{\mathrm{med}}
\newcommand{\Var}{\mbox{Var}}

\renewcommand{\thesection}{\Alph{section}}
\renewcommand{\thetable}{A.\arabic{table}}

\setcitestyle{authoryear,open={(},close={)}}

\baselineskip=24pt
\begin{center}
{\Large \bf
Supporting information for
``Approximate Bayesian Computation Design for Phase I Clinical Trials''\\
by Huaqing Jin, Wenbin Du and Guosheng Yin
}
\end{center}

\vspace{2mm}
\section{Simulation Details}
\subsection{Detailed settings of compared methods}
The detailed settings of the BOIN, CCD, CRM, keyboard, mTPI and UMPBI methods used in the simulation studies are listed as follows. 
\begin{itemize}
   \item {\bf BOIN:} In the BOIN design, we choose $\phi_1=0.6\phi$ and $\phi_2=1.4\phi$. Such setting follows \cite{lin2017nonparametric} and \cite{liu2015bayesian}.

   \item {\bf CCD:} Following \cite{ivanova2007cumulative}, we set the tolerance interval of the CCD method as $(0.2, 0.4)$ when $\phi=0.3$ and $(\phi-0.09, \phi+0.09)$ when $\phi < 0.3$.

    \item {\bf CRM:} we adopt the power model 
   $p_j = \pi_j^{\exp(\alpha)}$ with the model skeleton selected by the method of \cite{lee2009model}. 
   We choose a halfwidth of the indifference interval of $0.05$ and an initial guess of MTD at dose level $\lceil K/2 \rceil$.
   Note that such choices are popular in the literature \citep{lin2017nonparametric,lin2018uniformly}.
   
   \item {\bf Keyboard:} Following \cite{yan2017keyboard}, we set the proper dosing interval as $(\phi-0.05, \phi+0.05)$ for the keyboard design.
   
   \item {\bf mTPI:} Following the discussion in \cite{ji2010modified}, we choose the equivalent interval as $(\phi-0.05, \phi+0.05)$.
   
   \item {\bf UMPBI:} Following \cite{lin2018uniformly}, the threshold parameter, i.e., the only tuning parameter, is selected as 
  $\gamma(m_k) = \exp\left(c\sqrt{m_k}\right)$,
   with $c=\log(1.1)/3$.
\end{itemize}

For the real data application, we follow the original paper \citep{banerjee2017phase} and use the CRM design with a 2-parameter logistic model, i.e., 
\bse
{\rm logit}(p_j) = \alpha + \exp(\beta) x_i,
\ese
where the
model skeleton is still selected by the method of \cite{lee2009model} with  
   a halfwidth of the indifference interval of $0.05$ and an initial guess of MTD at dose level $\lceil K/2 \rceil$.
 The early stopping rule is set as terminating the trial if $\Pr(p_1>\phi|{\rm data}) > 0.95$.

\subsection{Random Scenario Generation}
We generate random scenarios to assess the performance of the phase I designs
in Sections 3.1 and 3.2 with the method of \cite{paoletti2004design}.
Specifically, the procedure is detailed as follows. 
\begin{enumerate}
    \item Randomly select, with equal probabilities,  one of the $K$ dose levels as the MTD and denote that dose level as $\wt{k}$.
    \item Let $\Phi$ be the cumulative density function (CDF) of the standard normal distribution.
    The probability of the MTD is $p_{\wt{k}}=\Phi(\epsilon_{\wt{k}})$ with $\epsilon_{\wt{k}} \sim N(\Phi^{-1}(\phi), \sigma_0^2)$, where $\phi$ is the target toxicity probability.
    \item For $\{p_k\}_{k=1}^{\wt{k}-1}$, generate 
    \bse
    p_{k-1} = \Phi\left[ 
    \Phi^{-1}(p_{k}) - 
    \left\{\Phi^{-1}(p_{k})-\Phi^{-1}(2\phi-p_{k})\right\}
    I\left\{\Phi^{-1}(p_{k})>\Phi^{-1}(\phi)\right\}
    - \epsilon_{k-1}^2
    \right],
    \ese
    where $I(\cdot)$ is the indicator function and 
    $\epsilon_{k-1} \sim N(\mu_1, \sigma_1^2)$.
    
    \item For $\{p_k\}_{k=\wt{k}+1}^{K}$, generate 
    \bse
    p_{k+1} = \Phi\left[ 
    \Phi^{-1}(p_{k}) + 
    \left\{\Phi^{-1}(2\phi-p_{k}) - \Phi^{-1}(p_{k})\right\}
    I\left\{\Phi^{-1}(p_{k})<\Phi^{-1}(\phi)\right\}
    + \epsilon_{k+1}^2
    \right],
    \ese where 
    $\epsilon_{k+1} \sim N(\mu_2, \sigma_2^2)$.
\end{enumerate}
Following \cite{liu2015bayesian}, 
we choose $\sigma_0 =0.05$ and $\sigma_1=\sigma_2=0.35$, 
and tune the parameters $\mu_1=\mu_2$ to achieve desirable $\Delta$, i.e., the average probability difference around the target.

\newpage

\bibliographystyle{abbrvnat}
\bibliography{ref.bib}%